\documentclass[11pt]{article}
\pdfoutput=1
\usepackage{jheppub}
\usepackage{graphicx}
\usepackage{array,tikz,tikz-cd,float}
\usetikzlibrary{positioning, quotes, decorations.markings, decorations.pathmorphing, shapes, calc}
\usepackage{color}
\usepackage{bm, dsfont, braket, slashed}

\tikzset{middlearrow/.style={
		decoration={markings,
			mark= at position 0.5 with {\arrow{#1}} ,
		},
		postaction={decorate}
	}
}

\newcommand{\pd}{{\partial}}
\newcommand{\opd}{{\overline \partial}}

\newcommand{\wt}{\widetilde}

\newcommand{\n}{\mathfrak{n}}

\newcommand{\Tr}{\text{Tr}}

\newcommand{\cf}{\emph{cf.}}
\newcommand{\ie}{\emph{i.e.}}
\newcommand{\eg}{\emph{e.g.}}
\newcommand{\aka}{\emph{a.k.a.}}

\newcommand{\be}{\begin{equation}}
	\newcommand{\ee}{\end{equation}}
\newcommand{\bp}{\begin{pmatrix}}
	\newcommand{\ep}{\end{pmatrix}}
\newcommand{\bsp}{\left(\begin{smallmatrix}}
	\newcommand{\esp}{\end{smallmatrix}\right)}

\newcommand*\diff{\mathop{}\!\mathrm{d}}

\newcommand{\D}{{\mathbb D}}
\newcommand{\R}{{\mathbb R}}

\newcommand{\C}{{\mathbb C}}
\newcommand{\Z}{{\mathbb Z}}

\newcommand{\CA}{{\mathcal A}}
\newcommand{\CB}{{\mathcal B}}
\newcommand{\CC}{{\mathcal C}}
\newcommand{\CD}{{\mathcal D}}

\newcommand{\CH}{{\mathcal H}}

\newcommand{\CM}{{\mathcal M}}
\newcommand{\CN}{{\mathcal N}}
\newcommand{\CO}{{\mathcal O}}

\newcommand{\CR}{{\mathcal R}}

\newcommand{\CT}{{\mathcal T}}
\newcommand{\CY}{{\mathcal Y}}

\newcommand{\CX}{{\mathcal X}}

\newcommand{\CL}{{\mathcal L}}

\DeclareMathOperator{\Lie}{Lie}
\DeclareMathOperator{\Spin}{Spin}

\newcommand{\fg}{\mathfrak{g}}

\newcommand{\ow}{{\overline{w}}}
\newcommand{\ox}{{\overline{x}}}
\newcommand{\oz}{{\overline{z}}}

\newcommand{\oQ}{{\overline{Q}}}
\newcommand{\oX}{{\overline{X}}}
\newcommand{\oY}{{\overline{Y}}}

\newcommand{\otheta}{{\overline{\theta}}}

\newcommand{\omu}{{\overline{\mu}}}
\newcommand{\onu}{{\overline{\nu}}}

\newcommand{\bA}{\mathbf{A}}
\newcommand{\bB}{\mathbf{B}}

\newcommand{\bO}{\mathbf{O}}

\newcommand{\bS}{\mathbf{S}}
\newcommand{\bT}{\mathbf{T}}
\newcommand{\bW}{\mathbf{W}}
\newcommand{\bX}{\mathbf{X}}
\newcommand{\bY}{\mathbf{Y}}
\newcommand{\bZ}{\mathbf{Z}}
\newcommand{\bGamma}{\mathbf{\Gamma}}

\newcommand{\bPsi}{\mathbf{\Psi}}
\newcommand{\bPhi}{\mathbf{\Phi}}

\newcommand{\bLambda}{\mathbf{\Lambda}}
\newcommand{\bmu}{\boldsymbol{\mu}}

\newcommand{\bOmega}{\mathbf{\Omega}}

\newcommand{\Sb}{\textrm{Sb}}

\newcommand{\Fc}{\textrm{Fc}}

\newcommand{\dR}{\textrm{dR}}
\newcommand{\Dol}{\textrm{Dol}}
\newcommand{\Maps}{\textrm{Maps}}
\newcommand{\Sect}{\textrm{Sect}}
\newcommand{\Bun}{\textrm{Bun}}
\newcommand{\Loc}{\textrm{Loc}}
\newcommand{\norm}[1]{{{:\!{#1}\!:}}}

\numberwithin{equation}{section}
\numberwithin{figure}{section}
\numberwithin{table}{section}

\title{Twisted Formalism for 3d $\CN=4$ Theories}
\author[1]{Niklas Garner}

\affiliation[1]{Department of Physics, University of Washington, Seattle, WA 98195}

\emailAdd{nkgarner@uw.edu}

\abstract{We describe the topological $A$ and $B$ twists of 3d $\CN=4$ theories of hypermultiplets gauged by $\CN=4$ vector multiplets as certain deformations of the holomorphic-topological ($HT$) twist of those theories, utilizing the twisted superfields of Aganagic-Costello-Vafa-McNamara describing $HT$-twisted 3d $\CN=2$ theories. We rederive many known results from this perspective, including state spaces on Riemann surfaces, deformations induced by flavor symmetries, the boundary VOAs of Costello-Gaiotto, and the category of line operators as proposed by Costello-Dimofte-Gaiotto-Hilburn-Yoo. Along the way, we show how the secondary product of local operators in the holomorphic-topological twist is related to the secondary product in the fully topological twist.}

\begin{document}
\today
\maketitle

\section{Introduction}
Recent studies have shown that mixed holomorphic-topological field theories spanning various dimensions possess quite rich structures, see \eg\, \cite{WittenHT, KapchiraldR, Witten2d02, AganagicCostelloMcNamaraVafa, GwilliamWilliams, CostelloDimofteGaiotto-boundary, CostelloNotes, OhYagi, Kapustin-hol, Costello-Yangian, EllYooLanglands} for examples arising from twisting supersymmetric theories. In special cases, these mixed holomorphic-topological theories can be deformed to fully topological theories, and such a realization can highly constrain the structure of the resulting topological theory. The prototypical example of this phenomenon is the deformation of mixed $BF$ theory to Chern-Simons theory, \cf\, \cite{AganagicCostelloMcNamaraVafa, GwilliamWilliams}. Gwilliam-Williams used this deformation to show that there is a choice of gauge in Chern-Simons theories that is manifestly 1-loop exact (when restricted to 3-manifolds equipped with a transverse holomorphic foliation); indeed, they show the vastly more general result that any ``chiral" deformation of the underlying mixed $BF$ theory admits such a choice of gauge \cite{GwilliamWilliams}. In particular, their analysis applies to the holomorphic-topological ($HT$) twist of a large class of $\CN=2$ Chern-Simons-matter theories, recently studied in \cite{CostelloDimofteGaiotto-boundary}. 

In this paper, we are interested in twisting 3d theories with $\CN=4$ supersymmetry. Three-dimensional theories with $\CN=4$ supersymmetry admit nilpotent supercharges that come in three types: up to spacetime symmetries and $R$-symmetry rotations, there is a single holomorphic-topological ($HT$) supercharge $Q_{HT}$, originally introduced for $\CN=2$ theories in \cite{CDFK1,CDFK2}, see also \cite{AOenvelopes, Butson2} for two recent appearances in $\CN=4$, and two distinct topological supercharges $Q_A$, a dimensional reduction of Witten's 4d Donaldson twist \cite{WittenTQFT}, and $Q_B$, an intrinsically 3d twist leading to Rozansky-Witten theory \cite{RW} for $\sigma$-models but originally studied by Blau-Thompson for pure gauge theory \cite{BT}. Importantly, the $A$ and $B$ topological supercharges $Q_A, Q_B$ are deformations of the $HT$ supercharge $Q_{HT}$.

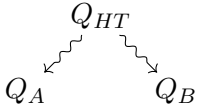
\begin{figure}[H]
	\centering
	\begin{tikzpicture}
		\draw (0,1) node {$Q_{HT}$};
		\draw (-1,0) node {$Q_A$};
		\draw (1,0) node {$Q_B$};
		\draw[->,decorate,decoration={snake,amplitude=.4mm,segment length=2mm,post length=1mm}] (-0.25,0.75) -- (-0.75,0.25);
		\draw[->,decorate,decoration={snake,amplitude=.4mm,segment length=2mm,post length=1mm}] (0.25,0.75) -- (0.75,0.25);
	\end{tikzpicture}
\end{figure}

The existence of deformations from $Q_{HT}$ to $Q_A$ and $Q_B$ implies that $HT$-twisted $\CN=4$ theories should admit two topological deformations, one to the $A$ twist and another to the $B$ twist.%
\footnote{Somewhat more precisely, if we write $Q_{A} = Q_{HT} + \delta_{A}$ then it is immediate that $\delta_A$ anticommutes with $Q_{HT}$ and is nilpotent. Thus, $\delta_A$ descends to a nilpotent symmetry of the $HT$-twisted theory. The $A$ twist, \ie\, twisting with $Q_A$, can then be realized by a spectral sequence whose first page corresponds to taking $HT$ twist, whose second page corresponds to taking $\delta_A$ cohomology of the $HT$-twisted theory, and so on. There are similar considerations for the $B$-twist.

In the present paper, we don't quite take this approach but use it as inspiration. Instead of taking the $HT$ twist, we consider a yet different theory (roughly obtained by formulating the theory in the Batalin-Vilkovisky formalism and integrating out $Q_{HT}$-exact fields) that is equivalent to the $HT$-twisted theory at the level of cohomology, \ie\, they are quasi-isomorphic. The deformations describe here are obtained by passing  $\delta_A, \delta_B$ through this quasi-isomorphism; importantly, we deform the theory at chain level, \ie\, we consider the total complex, rather than a spectral sequence.} %
For example, the algebra of local operators in the $HT$ twist should admit deformations to the algebras in the $A$ and $B$ twists. The algebra of local operators for a 3d $\CN=4$ theory in the $HT$ twist has the structure of a commutative vertex algebra (\ie\, OPEs of $Q_{HT}$-closed operators are always regular up to $Q_{HT}$-exact terms), just like the algebra of local operators in a $HT$-twisted $\CN=2$ theory. These $HT$-twisted algebras were described perturbatively by Oh-Yagi for free chiral multiplets \cite{OhYagi} and for more general $\CN=2$ gauge theories by Costello-Dimofte-Gaiotto \cite{CostelloDimofteGaiotto-boundary}. Aspects of the full, non-perturbative algebra of local operators was recently described by Zeng for abelian theories \cite{Zeng} but is not well understood for more general theories.

We take a more global perspective in this paper: we deform the \emph{entire} $HT$-twisted theory to a topological theory, \ie\, we consider $HT$-twisted 3d $\CN=4$ theories (viewed as $HT$-twisted $\CN=2$ theories) and deform them to their $A$ and $B$ twists in exact analogy with the deformation of mixed $BF$ theory to Chern-Simons theory. This perspective was emphasized by Elliott-Yoo in \cite{EllYooLanglands} for 4d $\CN=4$ super Yang-Mills, where the various Langlands (or Kapustin-Witten) twists \cite{KapustinWitten} are realized as deformations of a holomorphic twist.%
\footnote{See also the work of Elliott-Gwilliam-Williams \cite{EGW} for a similar discussion of the other twists of 4d $\CN=4$.} %
Recent work of Elliott-Safronov-Williams \cite{ESWtax} recast the $HT$, $A$, and $B$ twists for 3d $\CN=4$ super Yang-Mills theories using the BV-BRST formalism at the classical level; the above deformations were described by Butson from the perspective of factorization algebras \cite{Butson2}; and classical aspects of the holomorphic boundary conditions we consider in the present paper, including a refined analysis of the notion of ``deformable boundary condition'' from \cite{CostelloGaiotto}, were recently described by Brunner-Lavdas-Saberi \cite{BLS}. We find agreement whenever comparisons are possible.

As mentioned above, the fact that the $A$ and $B$ twists arise as deformations of an $HT$-twisted theory places strong constraints on the $A$ and $B$ twists, and it is these constraints that we use extensively in this paper. One salient consequence of the present paper, coupled with the analysis of \cite{GwilliamWilliams}, is that these $A$- and $B$-twisted theories (on any 3d manifold with a transverse holomorphic foliation) admit a choice of gauge that is manifestly 1-loop exact. We exploit this 1-loop exactness to perform exact, perturbative computations to derive the (perturbative sector of the) algebra of local operators on the holomorphic boundary conditions of \cite{CostelloGaiotto} and \cite{BLS}, described in more detail below. We also leverage the analysis of \cite{GwilliamWilliams} to show that there is no perturbative anomaly to the classical $Q$-exactness of the stress tensor, \ie\, these theories retain their topological nature through perturbation theory, and extend this result to a wide class of Chern-Simons-matter theories in the companion paper \cite{topCSM}.

In the recent paper \cite{CDGG}, we propose a physical QFT underlying the abstract 3d TQFTs of \cite{BCGPM, BGPMR} based on the quantum group $U_q(\mathfrak{sl}(n))$, providing several non-trivial checks of our proposal from various perspectives.  The present paper was inspired in part by a desire to explicate one of the central tools used in \cite{CDGG}, namely deforming an $HT$-twisted $\CN=4$ theory to its topological $A$ and $B$ twists. In the companion paper \cite{topCSM}, we use the same idea to twist the exotic $\CN=4$ Chern-Simons-matter theories of Gaiotto-Witten \cite{GaiottoWitten-Janus} and thereby propose boundary VOAs that optimistically encode many aspects of the bulk TQFT, akin to the WZW model in the classic Chern-Simons/WZW correspondence \cite{WittenJones} as well as the VOAs of \cite{CostelloGaiotto} for the standard $\CN=4$ gauge theories described in this paper. The theories investigated in \cite{CDGG} are examples in this collection: they are the Gaiotto-Witten theories $T[SU(n)]/SU(n)_k$ obtained by gauging the $SU(n)$ flavor symmetry of the 3d $\CN=4$ $T[SU(n)]$ theory of \cite{GaiottoWitten-Sduality} at a non-zero Chern-Simons level $k$. In particular, the results of the present paper and the companion paper \cite{topCSM} show that the topological nature of the $A$-twisted theories presented in \cite{CDGG} does not suffer from a perturbative anomaly.

\subsection{Summary of results}
We now briefly summarize our results.

\subsubsection{Topological deformations to $HT$-twisted theories}
One of the main objectives of this paper is to describe some general aspects of deforming mixed holomorphic-topological 3d theories to fully topological theories in analogy with the deformation of mixed $BF$ to Chern-Simons theory utilized by \cite{GwilliamWilliams}. For the purpose of this paper, we restrict our attention to the relatively simple example of deforming $HT$-twisted 3d $\CN=4$ theories of hypermultiplets gauged with $\CN=4$ vector multiplets to their topological $A$ and $B$ twists; we consider free hypermultiplets in Section \ref{sec:hypers} and generalize to gauge theories in Section \ref{sec:SYM}. See the companion paper \cite{topCSM} for an analysis of the exotic 3d $\CN=4$ Chern-Simons-matter theories of Gaiotto and Witten \cite{GaiottoWitten-Janus}. In order to describe this deformation efficiently, we use the twisted formalism of \cite{CostelloDimofteGaiotto-boundary, AganagicCostelloMcNamaraVafa}. This formalism is an application of the BV-BRST formalism to 3d $\CN=2$ theories that combines the twisting supercharge with the BV-BRST supercharge; this allows for a simplification of the field content and repackaging in terms of ``twisted superfields" that make manifest the trivialization of the $\oz$- and $t$-dependence of operators.

An important feature of the twisted formalism is its compatibility with higher operations, \eg\, those obtained from descent \cite{WittenTQFT,descent}. For example, the $z$-dependence of operators is measured by such an operation: one performs a surface integral of the stress tensor $*(T_{z\mu} \diff x^\mu)$ as in the physical theory.  Aspects of this operation for $HT$-twisted $\CN=2$ theories are discussed in \cite[Section~2.2]{CostelloDimofteGaiotto-boundary}. A necessary and sufficient condition for our deformed theory to be topological is for this operation to be cohomologically trivial, \ie\, operators \emph{do not} depend on $z$ in $Q$-cohomology; we explain how this operation trivializes upon deforming to a topological twist in Section \ref{sec:Qstress} and prove that all of the deformations considered in this paper, as well as those introduced in the companion paper \cite{topCSM}, are quantum mechanically topological, at least perturbatively.

Another operation of interest is the secondary product of local operators. $HT$-twisted theories admit a (degree $-1$) odd bracket $\{\!\{-, -\}\!\}_{HT}$ obtained from a holomorphic-topological decent procedure \cite{CostelloDimofteGaiotto-boundary, OhYagi}. The $HT$ descent bracket will be identically zero in cohomology after our topological deformations, indicating the presence of a yet-higher operation: an even bracket $\{\!\{-,-\}\!\}$ (of degree $-2$) coming from purely topological descent. The relation between the holomorphic-topological descent bracket $\{\!\{-,-\}\!\}_{HT}$ and the purely topological descent bracket $\{\!\{-,-\}\!\}$ is discussed in Section \ref{sec:descent} and we illustrate the procedure in our examples.

\subsubsection{Boundary VOAs, state spaces, line operators, and flavor backgrounds}
Twisted descriptions of the topological $A$ and $B$ twists of the gauge theories we consider are well understood, see \eg\, \cite{Butson2, CostelloPSI, KQZ, ESWtax}, but we consider their concrete and concise formulation in terms of the twisted superfields of \cite{AganagicCostelloMcNamaraVafa, CostelloDimofteGaiotto-boundary}. From this perspective, we provide straightforward derivations of many known results on boundary VOAs, BPS state spaces and line operators, and deformations induced by flavor symmetries.

We start with considering local operators bound to certain $\CN=(0,4)$ boundary conditions first proposed by \cite{CostelloGaiotto} to furnish a boundary VOA whose representation theory describes the category of line operators in the bulk, \cf\, the description of Wilson lines in Chern-Simons theory in terms of a chiral WZW model; a thorough description of classical aspects of the boundary conditions was recently given in \cite{BLS}. In Sections \ref{sec:bdyFHA} and \ref{sec:bdySYMA} (resp. \ref{sec:bdyFHB} and \ref{sec:bdySYMB}) we show quite explicitly that the $A$ (resp. $B$) twist VOAs proposed by \cite{CostelloGaiotto} lie on the boundary of our $A$-twisted (resp. $B$-twisted) theories via a straightforward perturbative analysis of the algebra of boundary local operators. One consequence of working in the twisted formalism is an absence of the subtleties requiring a suitable deformation of the desired boundary conditions in the physical theory, \cf\, \cite[Section 2]{CostelloGaiotto} or \cite[Section 3]{BLS}. It is important to note that the boundary conditions used in the $B$ twist are Dirichlet on the gauge fields and therefore admit boundary monopole operators, \cf\, \cite[Section 7]{CostelloDimofteGaiotto-boundary}, \cite[Section 6]{CostelloGaiotto}, and \cite[Section 3]{BDGH}, which are manifestly non-perturbative and beyond the reach of the present analysis.

In Sections \ref{sec:hilbFHA} and \ref{sec:hilbSYMA} (resp. Sections \ref{sec:hilbFHB} and \ref{sec:hilbSYMB}) we describe BPS state spaces $\CH_{A}(\Sigma)$ (resp. $\CH_{B}(\Sigma)$) on a Riemann surface $\Sigma$ in the $A$ and (resp. $B$) twist, matching expected results, \cf\, \cite{RW, BFKHilb, SafronovWilliams}.%
\footnote{As in \cite{CDGG}, the state spaces described in this paper are not quite the same as those in \cite{BFKHilb}, although they come from different choices of polarizations of the same phase space. In particular, the polarizations chosen in this paper has bounded cohomological degrees with finite-dimensional graded components, at the cost of lacking a Hermitian inner product, \cf\, \cite[Sec 2.5.3]{CDGG}. For the $A$-twisted theories, it is the same as the recent paper \cite{SafronovWilliams} describing BPS state spaces in $A$-twisted theories across various dimensions, although the authors of that paper use a slightly different twisting homomorphism than the one used in this paper -- they use the $U(1)_M$ flavor symmetry described below to make the field we call $X$ a scalar on $\Sigma$ rather than a spinor.} %
We obtain these results via geometric quantization of the equations of motion on $\Sigma$, \cf\, \cite{EllYooLanglands} or \cite{Costello02}. Using a state-operator correspondence, the state space for $\Sigma = S^2$ can be identified with the vector space of local operators and from our analysis we recover the mathematical description of the Coulomb branch due to Braverman-Finkelberg-Nakajima \cite{BFNII} in the $A$ twist as well as the Higgs branch in the $B$-twist. Using results of \cite{SafronovGQ} concerning geometric quantization in shifted symplectic geometry, we can extend this analysis to the next categorical level, \ie\, to the category of line operators, by considering the equations of motion on the (formal) punctured disk $\D^\times$, which is the algebraic version of the $S^1$ link of a line in 3d. The mathematical categories underlying the BPS line operators in question originate from unpublished work of Costello-Dimofte-Gaiotto-Hilburn-Yoo but are stated explicitly in, \eg\,, \cite[Section 1.1]{linevortex}, \cite[Section 7.7]{BFcoulomb}, or \cite[Section 1.2.8]{HilburnRaskin}; see also \cite{Hilburn, Yoo}.

Finally, in each example we describe a class of deformations that arise from turning on background fields associated to flavor symmetries. We focus on symmetries that are visible in the action. Namely, Higgs branch flavor symmetries $G_H$, to which background vector multiplets couple, and a maximal torus $T_C$ of the Coulomb branch flavor symmetry $G_C$, to which background twisted vector multiplets couple.%
\footnote{In the topological $A$ twist, vector multiplets localize to holomorphic bundles (or, more generally, monopole configurations) while twisted vector multiplets localize to bundles equipped with flat connections; these roles are exchanged in the $B$ twist. Thus, the $A$-twisted (resp. $B$-twisted) theory can deformed by background holomorphic $G_H$  (resp. $G_C$) bundles and $G_C$ (resp. $G_H$) bundles with flat connection.} %
Deformations by these flavor symmetry backgrounds played a central role in recent the partition function analysis of \cite{GHNPPS} and state space analysis of \cite{BFKHilb}. Additionally, the above supersymmetric state spaces form sheaves over the moduli space of such background bundles \cite{GaiottoTwisted}. More generally, the entire category of bulk line operators $\CC$ should admit deformations $\CC \rightsquigarrow \CC_\CA$ by background flat connections $\CA$ coupling to these flavor symmetries, where $\CC_\CA$ denotes the category of line operators compatible with the background $\CA$. The work \cite{CDGG} relates this deformation to the notion of a relative modular category \cite{DRnonsemisimple} appearing in recent mathematical works on 3d TQFTs based on quantum groups, \eg\, \cite{BGPMRholonomy, BCGPM, DRGPM}.

It is also important to note that these deformations are crucial ingredients in coupling twisted 3d $\CN=4$ theories to twisted 4d $\CN=4$ Yang-Mills. Indeed, if a 3d $\CN=4$ theory has a (Higgs or Coulomb) flavor symmetry then it can be used to engineer a boundary condition of 4d $\CN=4$ by gauging this flavor symmetry with the bulk 4d fields. These boundary conditions are compatible with taking the $A$ and $B$ twists in the combined bulk/boundary system have a remarkably rich interplay with $S$-duality in 4d and mirror symmetry in 3d. They thus play a central role in the Geometric Langlands program; see, \eg\, \cite{GaiottoLanglands} for physical aspects and \cite{HilburnRaskin} for a mathematically rigorous analysis in the simplest, non-trivial mirror pair of a hypermultiplet and a $U(1)$-gauged hypermultiplet. Forthcoming work of Ben-Zvi, Sakellaridis, and Venkatesh \cite{BZWHCGP} uses these boundary conditions to great extent by drawing upon profound analogies with number theory and harmonic analysis.

\subsection{Future directions}
There are many interesting future directions and applications of this work, some of which are the following.

\subsubsection{$\Omega$ backgrounds}
One aspect of 3d TQFTs that arise from twisting an underlying supersymmetric theory is the ability to quantize the secondary product of local operators \cite{descent} by working equivariantly with respect to rotations of spacetime around a fixed axis, thereby localizing the theory to an effective 1d theory along the axis. In the context of 3d $\CN=4$ theories, this is realized by turning on a 3d $\Omega$ background \cite{Yagi,BDG,BDGH}, which is a dimensional reduction of the 4d Nekrasov-Shatashvili $\Omega$ background \cite{NSomega}.

Given the twisted theories provided below, it is straightforward to incorporate an $\Omega$ deformation. This can be done by treating the 3d twisted theories as 1d twisted theories, as described for line operators above, where the $\Omega$ background corresponds to turning on a 1d twisted mass for the symmetry \cite{BDGHK}. For example, for an $A$-twisted free hypermultiplet one would consider a twisted action of the form 
\be
	S = S_A + \epsilon \int \iota_V \bY \pd \bX\,,
\ee
where $\iota_V$ denotes contraction with the holomorphic vector field $V = z \pd_z$ generating (complexified) rotations and $\epsilon$ is the complex mass/equivariant parameter of the $\Omega$ background.

\subsubsection{Boundary conditions}
The boundary conditions we consider in this paper are chiral, \ie\, boundary conditions that are compatible with 2d $\CN=(0,\bullet)$ supersymmetry and furnish chiral algebras of boundary local operators. The 3d $\CN=4$ theories we describe below also admit many interesting topological $\CN=(2,2)$ boundary conditions that appear in constructing modules for ($\Omega$-background deformed) Higgs and Coulomb branches \cite{BDGH, BDGHK, HKW} and their line operator generalizations \cite{linevortex, GK}. These modules are realized by the bulk-boundary map discussed in \cite{CostelloDimofteGaiotto-boundary} applied to the $A$- and $B$-twisted theories below.

\section{Twisted 3d $\CN = 2$ Theories}
\label{sec:N=2}
In this section we review and introduce the essential ingredients of our approach to deforming the holomorphic-topological ($HT$) twist of 3d $\CN = 2$ theories to topological theories. We start with a lightning review of the language of twisted superfields for $HT$-twisted 3d $\CN=2$ theories \cite{CostelloDimofteGaiotto-boundary, AganagicCostelloMcNamaraVafa} in Section \ref{sec:3dtwisted}. Our main source of examples will come from honest 3d $\CN=4$ theories, with the topological deformations corresponding to the deformation from the $HT$ twist to the $A$ and $B$ twists. In Section \ref{sec:N=4asN=2} we discuss how to write $\CN=4$ theories as $\CN=2$ theories and the relevant additional symmetries such a theory possesses. Finally, in Section \ref{sec:descent} we discuss how a deformation from a $HT$-twisted theory to a topological theory can be used to relate holomorphic-topological descent to purely topological descent, focusing on the secondary product of local operators; we realize this process in explicit examples in later sections.

\subsection{Twisted formalism for 3d $\CN = 2$ theories}
\label{sec:3dtwisted}
We begin by discussing the essential features of the twisted formalism of \cite{AganagicCostelloMcNamaraVafa, CostelloDimofteGaiotto-boundary}. The utility of this twisted formalism is to dramatically simplify the field content of the theory without losing any of the derived structures admitted by local and extended operators, \eg\, higher operations obtained by descent.

\subsubsection{Twisting}
We start with the 3d $\CN=2$ supersymmetry algebra. On flat Euclidean $\R^3$, the 3d $\CN=2$ supersymmetry algebra has four generators $Q_\alpha, \oQ_\alpha$, $\alpha = \pm$, satisfying 
\be
\{Q_\alpha, \oQ_\beta\} = (\sigma^\mu)_{\alpha \beta} P_{\mu}\,,
\ee
where $(\sigma^\mu)^\alpha{}_\beta$ are the Pauli matrices. (Spinor indices are raised and lowered using the Levi-Civita symbol as $\chi_\alpha = \epsilon_{\alpha \beta} \chi^\beta$ and $\chi^\beta = \chi_\alpha \epsilon^{\alpha \beta}$, where $\epsilon_{+-} = \epsilon^{+-} = 1$.) This algebra admits a $U(1)_R$ $R$-symmetry with respect to which $Q_\alpha, \oQ_\alpha$ have charge $-1, 1$. Up to symmetries of the algebra, there is a unique nilpotent supercharge and hence twist \cite{EagerSaberiWalcher, EStwists}, which we take to be $Q_{HT} = \oQ_+$. If we write $\R^3 \cong \C_{z,\oz} \times \R_t$ as $z = x_1+i x_2, t = x_3$, the non-trivial anti-commutation relations involving $Q_{HT}$ are given by
\be
\{Q_{HT}, Q_\oz\} = P_{\oz} \qquad \{Q_{HT}, Q_t\} = P_t\,,
\ee
where $Q_\oz = \tfrac{1}{2} Q_+$ and $Q_t = -Q_-$. Thus, the cohomology of $Q_{HT}$ will behave holomorphically on $\C_{z,\oz}$ and topologically on $\R_t$, hence we call this a ``holomorphic-topological" ($HT$) twist, although it is often simply called the ``holomorphic twist."

The $Q_{HT}$ twist is compatible with spacetimes that locally look like $\C_{z,\oz} \times \R_t$ or $\C_{z,\oz} \times \R_{t \geq 0}$. More precisely, the $\CN=2$ theories we will be interested in preserve the full $U(1)_R$ $R$-symmetry. With only a $U(1)_R$ $R$-symmetry, it is not possible to define the $Q_{HT}$ twist on an arbitrary 3-manifold. Instead, we can work on a 3-manifold compatible with reduction of the Lorentz group to the subgroup $\Spin(2)_E \subset SU(2)_E$ preserving vectors tangent to $\C$. The interiors of such manifolds locally take the form $\C_{z,\oz} \times \R_t$ and transition functions between patches $\C_{z,\oz} \times \R_t$ and $\C_{z',\oz'} \times \R_{t'}$ are of the form
\be
z \to z'(z) \qquad \oz \to \oz'(\oz) \qquad t \to t'(z,\oz,t).
\ee
Boundaries are identical, but instead are modeled on $\C_{z,\oz} \times \R_{t\geq 0}$. Manifolds with this type of structure are said to have a transverse holomorphic foliation (THF).

With respect to the subgroup $\Spin(2)_E$, the supercharges $Q_\pm, \oQ_\pm$ have spin $J_0 = \pm \tfrac{1}{2}$. The twisting homomorphism, \cf\, \cite{WittenTQFT, EagerSaberiWalcher, EStwists}, simply amounts to working with respect to the ``twisted spin" $\Spin(2)_{E'}$ generated by $J$ given by
\be
	J = \tfrac{1}{2} R - J_0.
\ee
With this choice, the supercharge $Q_{HT}$ has twisted spin $J=0$ and $U(1)_R$ $R$-charge $R = 1$. Similarly, the supercharges $Q_\oz$ and $Q_t$ have $U(1)_R$ $R$-charge $R = -1$ and twisted spins $J=-1$ and $J=0$, respectively.

\subsubsection{BV-BRST and twisted superfields}
In the following sections we will be interested in $\CN=2$ theories of vector multiplets coupled to chiral multiplets. See, \eg\,, \cite{AHISS} for a review of the untwisted $\CN=2$ theories. Here, we focus on the description of $HT$-twisted $\CN=2$ Chern-Simons-matter theories in \cite{AganagicCostelloMcNamaraVafa, CostelloDimofteGaiotto-boundary}, which uses the Batalin-Vilkovisky (BV) formalism \cite{BV}.

The data of such a 3d $\CN=2$ Chern-Simons-matter theory is a compact gauge group $G_c$, a unitary representation $V$ of $G_c$, a $G_c$-invariant superpotential $W: V \to \C$, and a collection of Chern-Simons levels $k$. We decompose the matter representation $V$ as $V = \bigoplus V^{(r)}$, where $V^{(r)}$ contains the fields of $R$-charge $r$. Denote by $\fg_c = \Lie(G_c)$ the (real) Lie algebra of $G_c$, $G$ the complexification of $G_c$, and $\fg = \Lie(G)$ its (complex) Lie algebra.

Using the notation of \cite{CostelloDimofteGaiotto-boundary}, we define $\bOmega^{\bullet, (j)} := C^\infty(\R^3)[\diff t, \diff \oz] \diff z^j$, where we treat $\diff t, \diff \oz$ as Grassmann odd variables and $\diff z$ as Grassmann even. Although we work locally, such a definition makes sense for any spacetime with THF. There is a natural product coming from the wedge product of forms $\bOmega^{i,(j)} \otimes \bOmega^{i',(j')} \to \bOmega^{i+i',(j+j')}$ as well as an integration map $\int: \bOmega^{2,(1)}  \to \C$. There are two natural derivations on this algebra given by $\diff' = \pd_t \diff t + \pd_{\oz} \diff \oz$ and $\pd = \pd_z \diff z$, the former being an cohomological degree 1 (odd) derivation of twisted spin $J=0$ and the latter being a cohomological degree 0 (even) derivation of twisted spin $J=1$.

The twisted formulation of this class of 3d $\CN=2$ theories includes the following fields:
\begin{itemize}
	\item two components of the gauge field organized into the fermionic field $$A = A_t \diff t + A_\oz \diff \oz \in \bOmega^{1, (0)} \otimes \mathfrak{g},$$ with $A_t$ complexified by the real scalar $\sigma$ of the $\CN=2$ vector multiplet
	\item a coadjoint-valued bosonic field $$B = B_z \diff z \in \bOmega^{0,(1)} \otimes \mathfrak{g}^*,$$ identified in the physical theory with the curvature $\tfrac{1}{g^2} F_{zt}$ up to Chern-Simons terms
	\item a $V$-valued bosonic field $$\phi = \sum\limits_r \phi_{r} \diff z^{r/2} \in \bigoplus\limits_{r} \bOmega^{0,(r/2)} \otimes V^{(r)},$$ identified with the bosons in the chiral superfields after applying the twisting homomorphism turning $\Spin(2)_E$ scalars of $R$-charge $R = r$ to sections of $K_\C^{r/2}$
	\item a $V^*$-valued fermionic field $$\eta = \sum\limits_r \big(\eta_{r,t} \diff t + \eta_{r,\oz} \diff \oz \big) \diff z^{1-r/2} \in \bigoplus\limits_{r} \bOmega^{1,(1-r/2)} \otimes (V^{(r)})^*,$$ whose components are identified with the covariant derivatives of the conjugate scalar $\eta_t \sim \overline{D_\oz \phi}, \eta_\oz \sim \overline{D_t \phi}$
\end{itemize}
The components of $A$ and $B$ have $R$-charge $0$, $\phi_r$ has $R$-charge $r$, and $\eta_r$ has $R$-charge $-r$. In the BV formalism, we further include anti-fields $A^*, B^*, \phi^*, \eta^*$ for the fields $A, B, \phi, \eta$ and a differential $Q_{BV}$ schematically given by
\be
	Q_{BV}(\text{anti-field}) = \text{EOM for field} \qquad Q_{BV}(\text{field}) = \text{EOM for anti-field}\,.
\ee
The action for our twisted theory takes the form
\be
\label{eq:twistedS}
S = \int B F'(A) + \eta \diff'_A \phi + \tfrac{1}{2} (\eta^*)^2 \pd^2 W + \tfrac{k}{4\pi}\Tr(A \pd A),
\ee
where $\diff'_A = \diff' + A = (\pd_t + A_t) \diff t + (\pd_{\oz} + A_\oz) \diff \oz$ is the covariant derivative, $F'(A) = \diff'A + A^2$ is the corresponding curvature. An even better description of the action of $Q_{BV}$ uses the (shifted-)Poisson bracket $\{-,-\}_{BV}$ on the space of fields, called the $BV$-bracket, that pairs fields and anti-fields as
\be
\{\text{field}, \text{anti-field}\}_{BV} = \delta^{(3)} \diff {\rm Vol}\,   
\ee
from which one identifies $Q_{BV} = \{-,S\}_{BV}$. It is straightforward to derive the action of $Q_{BV}$ from either of these descriptions. For example, the fields transform as
\be
\begin{aligned}
	Q_{BV} A & = 0 & \qquad Q_{BV} B & = 0\\
	Q_{BV} \phi & = 0 & \qquad Q_{BV} \eta & = \eta^* \pd^2 W\\
\end{aligned}
\ee
because the only anti-field that appears explicitly in the action is $\eta^*$.

This action has two types of redundancies, which the BV formalism accounts for by including ghost fields (and their corresponding anti-fields, \aka\, anti-ghosts). The first is a familiar gauge redundancy, for which we introduce the usual BRST ghost $c$ (a $\mathfrak{g}$-valued, fermionic scalar with $R$-charge $R = 0$: $c \in \bOmega^{0,(0)} \otimes \mathfrak{g}$), under which the fields transform as
\be
\begin{aligned}
	\delta_c A & = \diff'_A c & \qquad  \delta_c B & = c\cdot B + \tfrac{k}{2\pi} \pd c\\
	\delta_c \phi & = c\cdot \phi & \qquad \delta_c \eta & = c\cdot \eta
\end{aligned}\,,
\ee
where $c \, \cdot$ denotes the infinitesimal action of $\mathfrak{g}$ with parameter $c$. The unusual variation of $B$ ensures that (for nonzero level $k$) $\tfrac{2\pi}{k} B_z, A_\oz, A_t$ transform as components of a full gauge field $\CA$ that transforms as $\delta_c \CA = \diff_\CA c$, \cf\, \cite{AganagicCostelloMcNamaraVafa}.

The second redundancy comes from the twisting supercharge itself. In particular, we introduce a $V^*$-valued, fermionic ghost $\psi = \sum_r \psi_r \diff z^{1-r/2} \in \bigoplus_r \bOmega^{0,(1-r/2)} \otimes (V^{(r)})^*$ and transformations
\be
\begin{aligned}
	\delta_\psi A & = 0 & \qquad \delta_\psi B & = -\mu\\
	\delta_\psi \phi & = 0 & \qquad \delta_\psi \eta & = \diff'_A \psi + \pd^2 W \eta
\end{aligned}\,,
\ee
where $\mu$ is the moment map for the $\mathfrak{g}$ action on the representation $T^*[1] V \cong V^*[1] \times V$; in components it reads $\mu_a = \psi_m (\tau_a)^m{}_n \phi^n$ where $\tau_a$ are the matrices representing the $\fg$ action on $V$. In the physical theory, $\psi$ can be identified with one of the fermions in the (anti-)chiral multiplets, after applying the twisting homomorphism.

There is a ghost number symmetry $U(1)_{\rm gh}$, under which the fields $A, B, \phi, \eta$ have ghost number ${\rm gh} = 0$, the anti-fields $A^*, B^*, \phi^*, \eta^*$ have ghost number ${\rm gh} = -1$, the ghosts $c, \psi$ have ghost number ${\rm gh} = 1$, and the anti-ghosts $c^*, \psi^*$ have ghost number ${\rm gh} = -2$. We define the cohomological grading $U(1)_C$ as the sum of $R$-charge and ghost number:
\be
	C = R + {\rm gh}.
\ee
It is also important to note that we are free to redefine the cohomological grading $C$ by mixing with other abelian symmetries of the theory; we will make use of this freedom below. The twisted theory is thus graded by parity (fermionic or bosonic), twisted spin (generated by $J$), and the cohomological grading (generated by $C$).\footnote{We work in conventions such that parity alone determines the graded-commutativity of observables. Indeed, the $R$-charge, and hence the cohomological grading, in the above class of $\CN=2$ theories may be non-integral.} Each of the variations $\delta_c$ and $\delta_{\psi}$ is fermionic and has cohomological grading $C = 1$ and twisted spin $J = 0$, as desired. We denote the total transformation by
\be
	Q_{HT} = Q_{BV} + \delta_c + \delta_\psi\,.
\ee

After introducing anti-fields, ghosts, and anti-ghosts, the above field theory can be concisely repackaged in terms of ``twisted superfields." Consider the transformations of $c, A$, and $B^*$; they are given by
\be
Q_{HT} c = c^2 \qquad Q_{HT} A = \diff'_A c \qquad Q_{HT} B^* = c \cdot B^* + F'(A)\,.
\ee
If we define $\bA = c + A + B^* \in \bOmega^{\bullet, (0)} \otimes \fg[1]$, where $[1]$ denotes a shift in cohomological degree by 1, these variations can be neatly repackaged as
\be
Q_{HT} \bA = F'(\bA) = \overset{\text{0-form}}{c^2} + \overset{\text{1-form}}{\diff'_A c} + \overset{\text{2-form}}{c \cdot B^* + F'(A)}.
\ee
We can similarly combine the remaining fields:
\be
\begin{aligned}
	\bB & = B + A^* + c^* \in \bOmega^{\bullet, (1)} \otimes \fg^*\\
	\bPhi_r & = \phi_r + \eta_r^* + \psi_r^* \in \bOmega^{\bullet, (r/2)} \otimes V^{(r)}\\
	\bPsi_r & = \psi_r + \eta_r + \phi_r^* \in \bOmega^{\bullet, (1-r/2)} \otimes (V^{(r)})^*[1]
\end{aligned}
\ee
The variation of these twisted superfields under $Q_{HT}$ will be given below in Eq. \eqref{eq:QHT}. The twisted spin $J$ and cohomological grading $C$ of these twisted superfields, as well as $\diff t, \diff \oz, \diff z,$ are collected in Table \ref{table:spincoho}.
\begin{table}[h!]
	\centering
	\begin{tabular}{c|c|c|c|c|c|c|c}
		& $\bA$   & $\bB$  & $\bPhi_r$      & $\bPsi_r$          & $\diff t$ & $\diff \oz$ & $\diff z$ \\ \hline
		$(J,C)$ & $(0,1)$ & $(1,0)$ & $(\tfrac{r}{2}, r)$ & $(1-\tfrac{r}{2}, 1-r)$ & $(0,1)$   & $(-1,1)$    & $(1,0)$ 
	\end{tabular}
	\caption{Twisted spin $J$ and cohomological grading $C$ of the twisted superfields and differential forms $\diff t, \diff \oz, \diff z$ in the holomorphic-topological twist. The twisted superfields $\bA,\bB$ come from an $\CN=2$ vector multiplet and $\bPhi, \bPsi$ come from an $\CN=2$ chiral multiplet with $R$-charge $r$.}
	\label{table:spincoho}
\end{table}

As mentioned above, the desired variations arise from the BV-bracket $\{\,-,-\,\}_{BV}$, which pairs fields and anti-fields, via $Q_{HT} = \{-, S\}_{BV}$. In terms of the twisted superfields, the BV-bracket is explicitly given by
\be
\label{eq:BVbracket}
	\{\bA(x),\bB(y)\}_{BV} = \{\bPhi(x), \bPsi(y)\}_{BV} = \delta^{(3)}(x-y) \diff {\rm Vol}\,.
\ee
The appropriate action can be neatly expressed in terms of the above twisted superfields as
\be
\label{eq:twistedaction}
S = \int \bB F'(\bA) + \bPsi \diff'_{\bA} \bPhi + \bW + \tfrac{k}{4 \pi} \Tr(\bA \pd \bA),
\ee
where $\bW = W(\bPhi)$. The $Q_{HT}$ variation of the twisted superfields is then given by
\be
\begin{aligned}
	\label{eq:QHT}
	Q_{HT} \bA & = \frac{\delta S}{\delta \bB} =  F'(\bA) \qquad & Q_{HT} \bB & = \frac{\delta S}{\delta \bA} =  \diff'_{\bA} \bB - \bmu + \tfrac{k}{2\pi} \pd \bA\\
	Q_{HT} \bPhi & = \frac{\delta S}{\delta \bPsi} = \diff'_{\bA} \bPhi \qquad & Q_{HT} \bPsi & = \frac{\delta S}{\delta \bPhi} = \diff'_{\bA} \bPsi + \frac{\pd \bW}{\pd \bPhi}
\end{aligned}
\ee
where $\bmu = \mu(\bPhi, \bPsi)$.

Nilpotence of $Q_{HT}$ and invariance of $S$ under $Q_{HT}$ are equivalent to $S$ solving the classical master equation \cite{CostelloDimofteGaiotto-boundary}. Moreover, $Q_{HT}^2 = 0$ {\em off-shell} by construction. The theories described by this twisting procedure are a (``chiral") deformation of the class of theories studied in \cite{GwilliamWilliams}. The results of \emph{loc. cit.} imply that these theories have consistent perturbative quantizations, \ie\,, these quantizations satisfy the quantum master equation (at any scale).%
\footnote{The paper \cite{GwilliamWilliams} uses the machinery of the homotopy RG flow of \cite{CostelloRGEFT} to make mathematically precise statements in perturbative theory. In this paper, we will almost entirely ignore these details and work at infinitely long length scales $L \to \infty$ to simplify the discussion.}

The twisted superfields described above help make holomorphic-topological descent manifest; if $\CO$ is the lowest component of a twisted superfield $\bO$ with $Q_{HT} \bO = \diff' \bO$, then $\CO$ is $Q_{HT}$-closed and the higher form components of $\bO$ describe the operators making $\pd_\oz \CO, \pd_t \CO$ cohomologically trivial. On the other hand, for a general $HT$-twisted theory, the operator $\pd_z \CO$ must be realized as a surface integral against the stress tensor:
\be
\pd_z \CO(z,\oz, t) = \oint_{S^2} *(T_{z \mu} \diff x^\mu) \CO(z,\oz,t)
\ee
In the twisted formalism, the stress tensor $*(T_{z \mu} \diff x^\mu)$ can be expressed as the 2-form component of the twisted superfield $\bT$, although one often needs to modify the na{\"i}ve stress tensor obtained via a N{\"o}ther procedure on the action in Eq. \eqref{eq:twistedaction}. Explicitly, the (classical) modified stress tensor in the above theories is given by
\be
	\bT = \iota_{\pd_z}\bigg(-\bB \pd \bA + \sum\limits_r\big((1-\tfrac{r}{2}) \bPsi_r \pd \bPhi_r - \tfrac{r}{2} \bPhi_r \pd \bPsi_r\big)\bigg)
\ee
This (modified) stress tensor is such that $Q_{HT} \bT$ is $\diff$-exact, \ie\,
\be
	Q_{HT} \bT = \diff \bT\,.
\ee
It is straightforward to see that the surface integral realization of $\pd_z$ is therefore $Q_{HT}$-closed due to Stokes' theorem \cite[Sec 2.2]{CostelloDimofteGaiotto-boundary}. More generally, we can replace $\pd_z$ by a general holomorphic vector field $V(z) \pd_z$ to get conserved currents $\bT_V$ that extend holomorphic translations to more general (holomorphic) changes of coordinates.

\subsection{3d $\CN = 4$ as 3d $\CN = 2$}
\label{sec:N=4asN=2}
The examples of topological deformations to $HT$-twisted theories discussed in this paper come from deforming the $HT$ twist of theories that admit $\CN=4$ supersymmetry. The 3d $\CN=4$ supersymmetry algebra has an $\Spin(4)_R \cong SU(2)_A \times SU(2)_B$ $R$-symmetry and is generated by 4 spinors $Q^{a \dot{a}}_\alpha$, where $a$ is an $SU(2)_A$ doublet index and $\dot{a}$ is an $SU(2)_B$ doublet index, with non-trivial anti-commutation relations given by
\be
	\{Q^{a \dot{a}}_\alpha, Q^{b \dot{b}}_\beta\} = \epsilon^{ab} \epsilon^{\dot{a} \dot{b}} (\sigma^\mu)_{\alpha \beta} P_{\mu}
\ee
in the absence of central charges. Up to symmetries of the algebra, there is a single holomorphic-topological supercharge $Q_{HT}$ and two topological supercharges $Q_A, Q_B$ given by
\be
	Q_{HT} = Q^{+ \dot{+}}_+ \qquad Q_A = \delta^\alpha{}_{a} Q^{a \dot{+}}_\alpha \qquad Q_B = \delta^\alpha{}_{\dot{a}} Q^{+ \dot{a}}_\alpha
\ee
As noted in the Introduction, the topological supercharge $Q_A$ (resp. $Q_B$) is given by deforming $Q_{HT}$ by $Q^{- \dot{+}}_-$ (resp. $Q^{+ \dot{-}}_-$).

If we denote $Q^1 = Q^{-\dot{-}},\, \oQ^1 = Q^{+\dot{+}},\, Q^2 = - Q^{-\dot{+}},$ and $\oQ^2 = Q^{+\dot{-}}$, the above supersymmetry algebra splits into two $\CN=2$ algebras
\be
	\{Q^i_\alpha, \oQ^j_\beta\} = \delta^{ij} (\sigma^\mu)_{\alpha \beta} P_{\mu}\,.
\ee
Since the subalgebra generated by $Q^1, \oQ^1$ contains the $HT$ supercharge $Q_{HT} = \oQ^1_+ = Q^{+\dot{+}}_+$, we will decompose our $\CN=4$ multiplets with respect to this subalgebra. One natural choice of $\mathcal{N}=2$ $R$-symmetry group is the diagonal torus $U(1)_R \hookrightarrow SU(2)_A \times SU(2)_B$, \ie\, $R = \tfrac{1}{2}(H_A + H_B)$, where $H_A, H_B$ are the generators of the diagonal tori in $SU(2)_A$ and $SU(2)_B$. The supercharges $Q^1, \oQ^1$ would then have charges $-1, 1$ while $Q^2, \oQ^2$ are both chargeless. Although this choice of is invariant under the $\Z_2$ mirror automorphism of the 3d $\mathcal{N}=4$ algebra that exchanges $a,b,\ldots \leftrightarrow \dot{a}, \dot{b}, \ldots$, it is not well-suited to global considerations.%
\footnote{We thank J. Hilburn for bringing this to our attention.} %
In particular, the resulting twisted spin $J = \tfrac{1}{2} R - J_0$ of the $HT$-twisted theory would take quarter-integral values and hence requires a choice of (possibly non-existent) fourth root of the canonical bundle $K_\Sigma^{1/4}$ on the ``spatial'' surface $\Sigma$. Phrased differently, the putative $R$-symmetry group $U(1)_R \hookrightarrow SU(2)_A \times SU(2)_B$ generated by $\tfrac{1}{2}(H_A + H_B)$ is $4\pi$ periodic; the properly normalized generator, \ie\, the minimal normalization so that $e^{2 \pi i R}$ acts as $1$, is $R = H_A + H_B$, but then $Q_{HT}$ has $R$-charge 2.

The next best thing is to consider the two $R$-symmetry groups separately, leading to what we call the $HT^A$-twist and the $HT^B$-twist, where the 3d $\CN=2$ $R$-charge is given as $R_{HT^A} = H_A$ and $R_{HT^B} = H_B$, respectively. Clearly, the $HT^A$ and $HT^B$ twists are thus exchanged with one another by the aforementioned $\Z_2$ mirror automorphism. Moreover, the $HT^A$ twist is naturally compatible with deforming to the topological $A$ twist, \ie\, the $R$-symmetry backgrounds required to put the $HT^A$ twist on a 3-manifold with THF agree with those needed for the topological $A$ twist on the same manifold, \ie\, the twisted spins match $J_{HT^A} = J_A$. Similar considerations apply to the $HT^B$ twist and its deformation to the topological $B$ twist.

\begin{figure}[H]
	\centering
	\begin{tikzpicture}
		\draw (-2,1.5) node {$\CT^{HT^A}$};
		\draw (-3,-0.25) node {$\CT^{A}$};
		\draw[->,decorate,decoration={snake,amplitude=.4mm,segment length=3mm, post length=1mm}] (-2.25,1.25) -- (-2.75,0.25);
		\draw (-4.5, 0.75) node {deform to $A$ twist};
		
		\draw (2.25,1.5) node {$\CT^{HT^B}$};
		\draw (2.75,-0.25) node {$\CT^B$};
		\draw[->,decorate,decoration={snake,amplitude=.4mm,segment length=3mm, post length=1mm}] (2,1.25) -- (2.5,0.25);
		\draw (4.25, 0.75) node {deform to $B$ twist};
		
		\draw[<->] (-1.25,1.5) -- (1.25,1.5);
		\draw (0,1.75) node {change twist};
		\draw (0,1.25) node {data};
	\end{tikzpicture}
	\caption{The two holomorphic-topological ($HT$) twists $HT^A$ and $HT^B$ naturally deform the topological $A$ and $B$ twist, respectively, and are related to one another by a change of twisting data. Three-dimensional mirror symmetry acts by exchanging a theory $\CT$ and its mirror $\CT^\vee$ as well as exchanging the left and right sides of this diagram.}
	\label{fig:HTAHTB}
\end{figure}
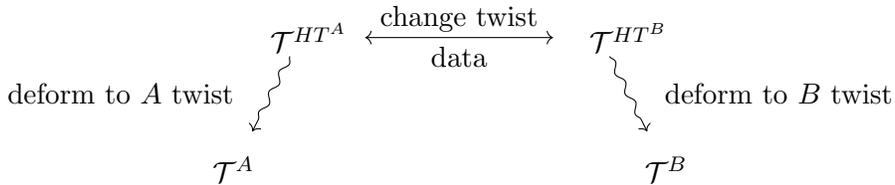

We now decompose the basic $\CN=4$ multiplets in terms of the above $\CN=2$ subalgebra. First, an $\CN=4$ hypermultiplet has two $SU(2)_A$ doublets of bosonic scalars $Z^{a n}$, $n=1,2$, and two $SU(2)_B$ doublets of fermionic spinors $\psi^{\dot{a} n}_\alpha$. Under the above $\CN=2$ subalgebra of the full $\CN=4$ algebra, this multiplet decomposes into two chiral/anti-chiral pairs: the bosonic components $X = Z^{+1}$, $Y = Z^{+2}$ fit into chiral multiplets with $\oX = Z^{-2}$, $\oY = -Z^{-1}$ as bosons of the corresponding anti-chiral partners. We will denote the twisted superfields coming from these chiral multiplets $(\bX, \bPsi_{\bX})$ and $(\bY, \bPsi_{\bY})$.

Using the above $R$-symmetry groups in the $HT^A$ and $HT^B$ twists, we conclude that in the $HT^A$ twist we have $r_X = r_Y = 1$ and in the $HT^B$ twist we have $r_X = r_Y = 0$. We write down the corresponding twisted spins in Table \ref{table:FHspin}.

\begin{table}[H]
	\centering
	\begin{tabular}{c|c|c|c|c}
		& $\bX$ & $\bPsi_{\bX}$ & $\bY$ & $\bPsi_{\bY}$ \\ \hline
		$J_A$ & $\tfrac{1}{2}$   & $\tfrac{1}{2}$    & $\tfrac{1}{2}$ & $\tfrac{1}{2}$\\ 
		$J_B$ & $0$   & $1$ & $0$ & $1$\\
	\end{tabular}
	\caption{Twisted spins $J_A, J_B$ for an $HT$-twisted hypermultiplet.}
	\label{table:FHspin}
\end{table}

The anti-diagonal torus $U(1)_F$, with $F = H_A - H_B$ acts trivially on $Q^1, \oQ^1$ and hence is a flavor symmetry from the $\mathcal{N}=2$ perspective. Additionally, there is an $SU(2)_M$ flavor symmetry rotating the $n$-index on the free hypermultiplet fields, and we find the pair of chiral multiplets transform as a doublet under this symmetry. The charges of our twisted superfields under the tori $U(1)_M$ and $U(1)_F$ are given in Table \ref{table:FHflavFM}.%
\footnote{It may happen that the torus $U(1)_M$ is part of the gauge group, hence not a flavor symmetry. Nonetheless, so long as the hypermultiplet representation is split $V \oplus V^*$ then this $U(1)_M$ action commutes with the gauge group and hence can be used to redefine the $U(1)_R$ charge.} %

\begin{table}[H]
	\centering
	\begin{tabular}{c|c|c|c|c}
		& $\bX$ & $\bPsi_{\bX}$ & $\bY$ & $\bPsi_{\bY}$ \\ \hline
		$F$ & $1$   & $-1$    & $1$ & $-1$\\
		$M$ & $1$   & $-1$    & $-1$ & $1$\\
	\end{tabular}
	\caption{Charges under the residual $R$-symmetry $U(1)_F$, viewed as a flavor symmetry of $\CN=4$ hypermultiplets, and under the diagonal torus $U(1)_M$ of the $SU(2)_M$ flavor symmetry rotating a free $\CN=4$ hypermultiplet.}
	\label{table:FHflavFM}
\end{table}

We will need these flavor symmetries to define the cohomological gradings in the $A$ and $B$ twists. In particular, the topological supercharge $Q_A$ is given by the sum of $Q_{HT} = \oQ^1_+$ and $-Q^2_-$. The twisted spin $J_A$ of $Q^2_-$ vanishes by design, but we will need to mix the cohomological grading with $U(1)_F$ charge to ensure that $Q_A$ has cohomological degree 1. Since $Q^2_-$ has charges $(0,-1)$ under $U(1)_{C_{HT^A}} \times U(1)_F$, the appropriate mixing corresponds to
\be
\label{eq:Amixing}
	C_A = C_{HT^A} - F = H_B + \mathrm{gh}\,.
\ee
It follows that both $\oQ^1_+$ and $-Q^2_-$ have twisted spin $J_A = 0$ and cohomological degree $C_A = 1$, as desired. We can apply the same analysis to the $B$-twist. It is conventional to use the maximal torus $U(1)_M \hookrightarrow SU(2)_M$ to ensure that bosonic fields in these chiral multiplets have even cohomological grading in the $B$ twist, \cf\, \cite[Sec 5.2.3]{descent}.%
\footnote{A similar phenomenon happens for monopole operators in the $A$ twist, where one should include a Coulomb-branch flavor symmetry in the cohomological grading. We will briefly comment on the need to do this in Section \ref{sec:hilbSYMA}.} %
We find that the appropriate mixing is given by
\be
\label{eq:Bmixing}
	C_B = C_{HT^B} + F - M = H_A + \mathrm{gh} - M\,.
\ee
We write down the cohomological degrees of the hypermultiplet twisted superfields in Table \ref{table:FHcoho}.
	
\begin{table}[h!]
	\centering
	\begin{tabular}{c|c|c|c|c}
		& $\bX$ & $\bPsi_{\bX}$ & $\bY$ & $\bPsi_{\bY}$ \\ \hline
		$C_A$ & $0$   & $1$    & $0$ & $1$\\
		$C_B$ & $0$   & $1$    & $2$ & $-1$\\
	\end{tabular}
	\caption{Cohomological degrees $C_A$ and $C_B$ of the twisted superfields for a hypermultiplet in the topological $A$ and $B$ twists.}
	\label{table:FHcoho}
\end{table}

Our theories will also have gauge fields in $\CN=4$ vector multiplets. In addition to the gauge fields, this supermultiplet has an $SU(2)_B$-triplet of real bosonic scalars $\phi^{(\dot{a} \dot{b})}$, an $SU(2)_A$-triplet of real bosonic auxiliary fields $D^{(ab)}$ (that are set equal to the moment maps $\mu^{ab}$ on-shell), and an $SU(2)_A \times SU(2)_B$ bi-doublet of fermions $\lambda^{a \dot{a}}_\alpha$.

Under our choice of $\CN= 2$ subalgebra of the full $\CN=4$ algebra, an $\CN = 4$ vector multiplet decomposes into an $\CN = 2$ vector multiplet and an chiral/anti-chiral pair with the chiral multiplet valued in the adjoint representation $\mathfrak{g}$. The $\CN=2$ vector multiplet is made up of the gauge field $A_\mu$, the fermions $\nu = \lambda^{+ \dot{-}}_\alpha, \onu = \lambda^{- \dot{+}}_\alpha$, the boson $\sigma = \phi^{(\dot{+}\dot{-})}$, and the auxiliary field $D = D^{(+-)}$. The $\CN=2$ chiral multiplet contains the boson $\phi = \phi^{(\dot{+}\dot{+})}$, the fermion $\lambda = \lambda^{-\dot{-}}_\alpha$, and the auxiliary field $F = D^{(++)}$, and similarly for the anti-chiral multiplet. We will denote the corresponding twisted superfields by $(\bA, \bB)$, for the $\CN=2$ vector multiplet, and $(\bPhi, \bLambda)$, for the $\CN=2$ chiral/anti-chiral pair. We write down the twisted spins and cohomological degrees of these twisted superfields in Table \ref{table:FVspincoho}.

\begin{table}[h!]
	\centering
	\begin{tabular}{c|c|c|c|c}
		& $\bA$   & $\bB$  & $\bPhi$ & $\bLambda$\\ \hline
		$(C_A, J_A)$ & $(1,0)$ & $(0,1)$ & $(0,2)$ & $(1,-1)$\\
		$(C_B, J_B)$ & $(1,0)$ & $(0,1)$ & $(0,1)$ & $(1,0)$\\
	\end{tabular}
	\caption{Twisted spins and cohomological degrees of the twisted superfields arising from a $\CN=4$ vector multiplet in the topological $A$ and $B$ twists.}
	\label{table:FVspincoho}
\end{table}

\subsection{Perturbative exactness of the stress tensor}
\label{sec:Qstress}
We will be interested in topological deformations of several examples of the above theories. Since the triviality of $\pd_t$ and $\pd_{\oz}$ is built into the formalism -- the higher form components of a twisted superfield are the holomorphic-topological descendants of the lowest component --  for such a deformation $Q_{HT} \rightsquigarrow Q$ to yield a topological theory, the generator of holomorphic translations $\pd_z$%
\footnote{We restrict to constant vector fields to simplify the discussion. A similar analysis to what follows can be applied to more general holomorphic vector fields $V$, showing that they too act trivially in $Q$ cohomology.} %
must be cohomologically trivial. In particular, it suffices to check that the stress tensor $\bT$ is $Q$-exact, up to terms proportional to $\diff$.

One of the advantages of working in the twisted formalism described above is that, under some very mild assumptions, the classical $Q$-exactness of $\bT$ \emph{implies} the perturbative%
\footnote{It is important to note that this need not be the case: in principle, there may be anomalies that cause the classical $Q$-exactness of the stress tensor to fail. Even the quantum existence of the stress tensor $\bT$ can fail; see, \eg\,, \cite{SilversteinWitten1} for a related discussion in the context of 2d $\CN=(0,2)$ theories. Although we expect it not to happen in the examples discussed in the present paper, there may be further non-perturbative obstructions, \cf\, \cite{SilversteinWitten2} for the corresponding statements in 2d $\CN=(0,2)$ theories. We leave the verification of the non-perturbative $Q$-exactness of $\bT$ in these and related examples to future work.} %
$Q$-exactness of $\bT$, as we now explain.%
\footnote{We thank B. Williams for explaining the following argument.} %
See \cite[Sec 5]{EGW} for a analogous discussion in the context of the Langlands/Kapustin-Witten twists 4d $\CN=4$ super Yang-Mills. Consider a theory of the form described in Section \ref{sec:3dtwisted}, or any chiral deformation thereof. More precisely, we require that the theory is a chiral deformation of the free mixed $BF$ theory $\int \bB \diff' \bA + \bPsi \diff' \bPhi$: the interactions are required to be polynomials in the twisted superfields and holomorphic derivative $\pd_z$. Suppose we have a (classical) local operator $\bS$ such that
\be
\label{eq:QST}
	Q\bS = \diff \bS + \bT\,.
\ee
Classically, the statement that $\bT$ generates holomorphic translations is encoded in the fact that $\{\int \bT, \bO\}_{BV} = \pd_z \bO$ for any $\bO$. It follows that $\{\int \bT,\int \bT\}_{BV} = 0 = \{\int \bT, \int \bS\}_{BV},$ each being the integral of a total derivative. For simplicity, we will further assume that $\{\int \bS, \int \bS\}_{BV} = 0$. Equation \eqref{eq:QST} implies $\bO^{\langle 1 \rangle} := \{\int\bS, \bO\}_{BV} \diff z$ satisfies
\be
\label{eq:antiholodescent}
	Q \bO^{\langle 1 \rangle} = (Q\bO)^{\langle 1 \rangle} + \pd \bO\,,
\ee
\ie\, $\bO^{\langle 1 \rangle}$ contains the $\diff z$ components of the various \emph{topological} descendants of (the lowest component of) $\bO$.

Phrased differently, the stress tensor $\bT$ realizes an action of the abelian Lie algebra $\C$ of holomorphic translations
in $z$ generated by the constant vector field $t = \pd_z$. The existence of the above $\bS$ then corresponds to the action of an extended DG Lie algebra $\C_{\dR}$ with the above degree 0 generator $t$, a degree $-1$ generator $s$, and differential $q$ given by $q s = t, q t = 0$. We think of $s$ a supertranslation generator, and the differential $q$ as the twisting supercharge. Note that there is an inclusion of DG Lie algebras $\C \hookrightarrow \C_{\dR}$; the data of an action of $\C_{\dR}$ on this classical theory, compatible with the action of $\C$ through this inclusion, is called a ``homotopy trivialization" of the $\C$ action, \cf\, \cite[Def. 5.2]{EGW}. If the classical action of $\C_{\dR}$ does not suffer from an anomaly, it follows that the quantization of $\bT$ generates holomorphic translations and, moreover, that the quantum analogues of Eq. \eqref{eq:QST} and Eq. \eqref{eq:antiholodescent} hold. Thus, it suffices to check this action is non-anomalous.

A straight-forward way to check that this symmetry is non-anomalous in the BV/BRST formalism is by coupling to background fields for the symmetry. In particular, we introduce two non-dynamical background fields $\tau, \sigma \in \C_{\dR}[1]$ and their anti-fields $\tau^*, \sigma^* \in \C_{\dR}^*$ and then deform the action as
\be
	S \to S' = S + \sigma^* \tau + \int \sigma \bS + \tau \bT\,,
\ee
so that the deformed BV/BRST supercharge $Q'$ acts on the background fields as
\be
\begin{aligned}
	Q' \tau & = 0 \qquad & Q' \tau^* & = \sigma^* + \int \bT\\
	Q' \sigma & = \tau \qquad & Q' \sigma^* & = -\int \bS
\end{aligned}
\ee
The action of $Q'$ on the background fields $\tau, \sigma$ encodes the action of the differential $q$ on the generators $t,s$; the variation of the anti-fields $\tau^*, \sigma^*$ is the current to which the symmetry couples (together with the induced action of $q$ on the dual space $\C_\dR^*$). In particular, the fact that $\C_{\dR}$ acts on the classical theory (via the currents currents $\bT, \bS$) translates to the statement that $S'$ solves the classical master equation, \cf\, \cite[Def. 13.2.2.1]{CGvol2} and \cite[Def. 2.16]{EGW}.

Finally, we appeal to \cite[Theorem 5.1]{GwilliamWilliams}: so long as $\int \bS$ and $\int \bT$ are chiral deformations, \eg\, if $\bS, \bT$ are polynomials in the fields and holomorphic derivatives $\pd_z$, there is no (perturbative) anomaly to quantizing the classical action $S'$. In particular, the $\C_{\dR}$ action, and hence the $Q$-exactness of the stress tensor $\bT$, persists in perturbation theory.

\subsection{Topological descent from holomorphic-topological descent}
\label{sec:descent}
As described in detail in \cite[Sec 3.4]{CostelloDimofteGaiotto-boundary}, there is a (degree $-1$) bracket $\{\!\{-,-\}\!\}_{HT}$ on the $Q_{HT}$ cohomology of local operators obtained by holomorphic-topological descent. When deforming the $HT$ twist to a topological theory, these operations become trivial at the level of cohomology. Nonetheless, there will be another secondary operation obtained by {\em topological} descent. We now relate the two operations, providing concrete examples in the subsequent sections. The following easily generalizes to other dimensions, where one hopes to deform purely holomorphic or mixed holomorphic-topological higher products to topological (or at least more topological) higher products.

Let $Q$ be the $HT$ supercharge, and suppose the theory possesses a nilpotent symmetry $\delta$ such that $\mathbb{Q} = Q + \delta$ is topological supercharge, in particular $\delta^2 = \{Q, \delta\} = 0$. Since $\mathbb{Q}$ is a topological supercharge, there are odd symmetry generators $Q_t, Q_z, Q_{\oz}$ such that%
\footnote{Somewhat more generally, the deformation $\delta$ need not be an exact symmetry of the $HT$ twist but still satisfies the Maurer-Cartan equation $\{Q, \delta\} + \delta^2 = 0$. Moreover, the elements $Q_t, Q_\oz$ trivializing $\pd_t, \pd_\oz$ in the $Q$ twist and the $\mathbb{Q}$ twist could be different. Neither of these situations arise in the examples considered in this paper, so we restrict to this simplified case. We expect that it is possible to generalize these arguments to more general cases.}
\be
\begin{aligned}
	\{\mathbb{Q}, Q_t\} = \{Q, Q_t\} = \pd_t & \qquad & \{\mathbb{Q}, Q_{\oz}\} = \{Q, Q_{\oz}\} = \pd_{\oz}\\
	\{\mathbb{Q}, Q_z\} = \pd_z & \qquad & \{Q, Q_z\} = 0\\
\end{aligned}.
\ee

Given an operator $\CO$ closed under $\mathbb{Q}$ and $Q$ (and hence $\delta$) we define the first descendants $\CO^{(1)}, \CO^{\langle 1 \rangle}$ by
\be
\CO^{(1)} = (Q_\oz \CO)\diff \oz + (Q_t \CO)\diff t \qquad \CO^{\langle 1 \rangle} = (Q_z \CO)\diff z
\ee
as well as $\CO^{[1]} = \CO^{(1)} + \CO^{\langle 1 \rangle}$. They satisfy the equations
\be
\mathbb{Q}\CO^{[1]} = \diff \CO \qquad \mathbb{Q} \CO^{(1)} = Q\CO^{(1)} = \diff' \CO \qquad \delta\CO^{\langle1\rangle} = \pd \CO.
\ee
We can similarly define the second descendants $\CO^{(2)}$ and $\CO^{[2]}$ as well as the mixed descendant $(\CO^{(1)})^{\langle 1 \rangle} = (\CO^{\langle 1 \rangle})^{(1)}$. They are related by $\CO^{[2]} = \CO^{(2)} + (\CO^{(1)})^{\langle 1 \rangle}$.

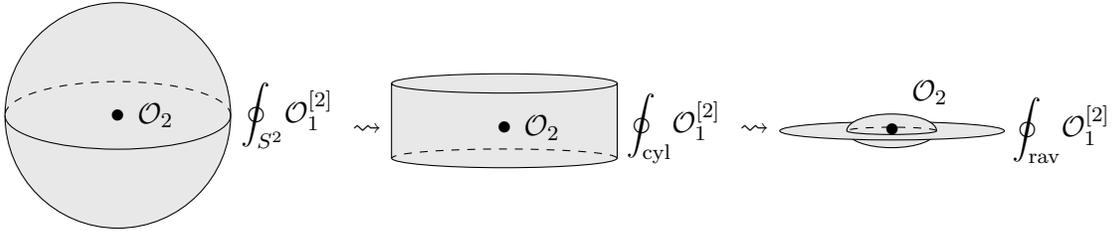
\begin{figure}[H]
	\centering
	\begin{tikzpicture}[scale=1]
		\draw [fill={rgb:black,1;white,10}] (0,0) circle (1.5cm);
		\draw (-1.5,0) arc (180:360:1.5 and 0.45);
		\draw[dashed] (1.5,0) arc (0:180:1.5 and 0.45);
		
		\draw (2.25,0) node {$\displaystyle{\oint_{S^2}} \CO_1^{[2]}$};
		\draw (0,0) node {$\bullet$};
		\draw (0.5,0) node {$\CO_2$};
	\end{tikzpicture}
	\raisebox{1.25cm}{$\rightsquigarrow$}
	\raisebox{0.75cm}{\begin{tikzpicture}[scale=1]
			\node [cylinder, shape border rotate = 90, aspect=1, draw=black, minimum height=1.25cm, minimum width=3cm,cylinder uses custom fill, cylinder body fill={rgb:black,1;white,10}, cylinder end fill={rgb:black,1;white,10}, alias=A] (0,0) {};
			\draw[dashed]
			let \p1 = ($ (A.after bottom) - (A.before bottom) $),
			\n1 = {0.5*veclen(\x1,\y1)-\pgflinewidth},
			\p2 = ($ (A.bottom) - (A.after bottom)!.5!(A.before bottom) $),
			\n2 = {veclen(\x2,\y2)-\pgflinewidth}
			in
			([xshift=-\pgflinewidth] A.before bottom) arc [start angle=0, delta angle=180,
			y radius=\n2, x radius=\n1];
			
			\draw (2.25,0) node {$\displaystyle{\oint_{\rm cyl}} \CO_1^{[2]}$};
			\draw (0,0) node {$\bullet$};
			\draw (0.5,0) node {$\CO_2$};
	\end{tikzpicture}}
	\raisebox{1.25cm}{$\rightsquigarrow$}
	\raisebox{0.75cm}{\begin{tikzpicture}[scale=1]
			\node [cylinder, shape border rotate = 90, aspect=1, draw=white, minimum height=2cm, minimum width=3cm] (A) {};
			
			\draw[fill={rgb:black,1;white,10}]
			let \p1 = ($ (A.after bottom) - (A.before bottom) $),
			\n1 = {0.5*veclen(\x1,\y1)-\pgflinewidth},
			\p2 = ($ (A.bottom) - (A.after bottom)!.5!(A.before bottom) $),
			\n2 = {veclen(\x2,\y2)-\pgflinewidth}
			in
			([xshift=-0.4*(\n1+\pgflinewidth)] $(A.bottom)$) arc [start angle=180, delta angle=180,
			y radius=0.15*\n1, x radius=0.4*\n1];
			
			\draw [black, fill={rgb:black,1;white,10}]
			let \p1 = ($ (A.after bottom) - (A.before bottom) $),
			\n1 = {0.5*veclen(\x1,\y1)-\pgflinewidth},
			\p2 = ($ (A.bottom) - (A.after bottom)!.5!(A.before bottom) $),
			\n2 = {veclen(\x2,\y2)-\pgflinewidth}
			in
			($(A.bottom)$) ellipse [y radius=\n2, x radius=\n1];
			
			\draw[fill={rgb:black,1;white,10}]
			let \p1 = ($ (A.after bottom) - (A.before bottom) $),
			\n1 = {0.5*veclen(\x1,\y1)-\pgflinewidth},
			\p2 = ($ (A.bottom) - (A.after bottom)!.5!(A.before bottom) $),
			\n2 = {veclen(\x2,\y2)-\pgflinewidth}
			in
			([xshift=-0.4*(\n1+\pgflinewidth)] $(A.bottom)$) arc [start angle=180, delta angle=180,
			y radius=0.4*\n2, x radius=0.4*\n1] arc [start angle=0, delta angle=180,
			y radius=0.15*\n1, x radius=0.4*\n1];
			
			\draw[dashed]
			let \p1 = ($ (A.after bottom) - (A.before bottom) $),
			\n1 = {0.5*veclen(\x1,\y1)-\pgflinewidth},
			\p2 = ($ (A.bottom) - (A.after bottom)!.5!(A.before bottom) $),
			\n2 = {veclen(\x2,\y2)-\pgflinewidth}
			in
			([xshift=0.4*(\n1-\pgflinewidth)] $(A.bottom)$) arc [start angle=0, delta angle=180,
			y radius=0.4*\n2, x radius=0.4*\n1];
			
			\draw (0,-0.9) node {$\bullet$};
			\draw (0.5,-0.4) node {$\CO_2$};
			
			\draw (2.25,-0.9) node {$\displaystyle{\oint_{\rm rav}} \CO_1^{[2]}$};
	\end{tikzpicture}}
	\caption{An illustration of the deformation of the surface integral defining the descent bracket $\{\!\{\CO_1, \CO_2\}\!\}$ from a sphere $S^2$ surrounding $\CO_2$ to a raviolo.}
	\label{fig:cyl2rav}
\end{figure}

We can use the relations between these descent procedures to connect the two types of secondary products. Suppose $\CO_1, \CO_2$ are $Q$-closed local operators that are invariant under $\delta$, and are therefore $\mathbb{Q}$-closed. The topological descent bracket between $\CO_1, \CO_2$ is given by the following surface integral
\be
\{\!\{\CO_1, \CO_2\}\!\} = \oint_{S^2} \CO_1^{[2]} \CO_2\,.
\ee
By Stokes' theorem, the value of this integral (in $\mathbb{Q}$-cohomology) only depends on the homology class of the surface. A particularly useful choice is a cylinder; in the limit that the height of the cylinder goes to zero, resulting in a ``raviolo"  as depicted in Figure \ref{fig:cyl2rav}, the only terms of $\CO_1^{[2]}$ that contribute to the integral are proportional to $\diff \oz \wedge \diff z$, in particular we have
\be
\{\!\{\CO_1, \CO_2\}\!\} = \oint_{{\rm rav}} \CO_1^{[2]} \CO_2 = \oint_{{\rm rav}} (\CO_1{}^{(1)}){}^{\langle 1 \rangle} \CO_2\,.
\ee

If $\wt{\CO}_1$ is a third $Q$-closed local operator, then the $HT$-descent bracket of $\wt{\CO}_1$ and $\CO_2$ is given by the surface integral%
\be
\{\!\{\wt{\CO}_1, \CO_2\}\!\}_{HT} := \oint_{S^2} (\wt{\CO}_1^{(1)}\wedge \diff z) \CO_2\,.
\ee
The value of this integral (in $Q$-cohomology) only depends on the homology class of the surface, and we can again choose it to be a raviolo. If we have $\wt{\CO}_1^{(1)} \wedge \diff z = (\CO^{(1)})^{\langle 1 \rangle}$, \eg\, by choosing $\wt{\CO}_1 = Q_z \CO_1$, it follows that the $HT$ descent bracket of $\wt{\CO}_1$ and $\CO_2$ agrees with the topological descent bracket of $\CO_1$ and $\CO_2$:
\be
	\{\!\{\wt{\CO}_1, \CO_2\}\!\}_{HT} = \oint_{{\rm rav}} (\wt{\CO}_1^{(1)}\wedge \diff z) \CO_2 = \oint_{{\rm rav}} (\CO_1{}^{(1)}){}^{\langle 1 \rangle} \CO_2 = \{\!\{\CO_1, \CO_2\}\!\}\,.
\ee

\section{Hypermultiplets}
\label{sec:hypers}	
In this section we consider the example of free $\CN=4$ hypermultiplets. We consider its $HT^A$ and $HT^B$ twist simultaneously and then deform the $HT^A$ twist (resp. $HT^B$ twist) to the $A$ twist (resp. $B$ twist) in Section \ref{sec:FHA} (resp. Section \ref{sec:FHB}).

As described in Section \ref{sec:N=4asN=2}, we will denote the twisted superfields coming from an $\CN=4$ hypermultiplet by $(\bX, \bPsi_{\bX})$ and $(\bY, \bPsi_{\bY})$, where the bottom components of the bosonic superfields $\bX$ and $\bY$ are the bosons $X$ and $Y$, respectively. Similarly, the bottom components of the fermionic twisted superfields $\bPsi_\bX$ (resp. $\bPsi_{\bY}$) $\psi_X$ (resp. $\psi_Y$) are fermions in the anti-chiral multiplets. Since there is no superpotential, the $HT$ twisted action is simply given by
\be
\label{eq:twistedactionFH}
	S = \int \bPsi_{\bX} \diff' \bX + \bPsi_{\bY} \diff' \bY.
\ee
The action of the $HT$($+$BV/BRST) supercharge $Q_{HT}$ is given by
\be
\begin{aligned}
	\label{eq:QFH}
	Q_{HT} \bX & = \diff' \bX \qquad & Q_{HT} \bPsi_{\bX} & = \diff' \bPsi_{\bX}\\
	Q_{HT} \bY & = \diff' \bY \qquad & Q_{HT} \bPsi_{\bY} & = \diff' \bPsi_{\bY}\\
\end{aligned}\,.
\ee
The algebra of local operators in this $HT$-twisted theory is simply the (commutative) vertex algebra generated by the bosons $X,Y$ and the fermions $\psi_X, \psi_Y$ \cite{OhYagi, CostelloDimofteGaiotto-boundary}. The non-trivial holomorphic-topological descent brackets on this commutative vertex algebra are given by
\be
\label{eq:descentFH}
	\{\!\{X,\psi_X\}\!\}_{HT} = \{\!\{Y,\psi_Y\}\!\}_{HT} = 1.
\ee

It is worth noting that the above twisted superfields can be combined into a more compact form as
\be
\label{eq:combinedhypers}
	\CX = \bX - \varepsilon \bPsi_{\bY} \qquad \CY = \bY + \varepsilon \bPsi_{\bX},
\ee
where $\varepsilon$ is an odd parameter of $U(1)_F \times U(1)_M$ charge $(2,0)$.%
\footnote{In the $HT^A$ twist, $\varepsilon$ is a scalar of cohomological degree $1$. In the $HT^B$ twist, it has spin $1$ (transforms as $\diff z$) and cohomological degree $-1$.} %
The fields $\bX, \bY$ are sections of $\Omega^{\bullet, (1/2)}$ over $\R \times \C$ in the $HT^A$ twist and sections of $\Omega^{\bullet, (0)}$ in the $HT^B$ twist. Similarly, the fields $\bPsi_\bX, \bPsi_{\bY}$ parameterize (odd, shifted) tangent vectors above these sections. Equivalently, the fields $\CX, \CY$ can be thought of as sections of $\Omega^{\bullet, (1/2)}$ or $\Omega^{\bullet, (0)}$ over $\R \times \C \times \C^{0|1}$. The equations of motion for these fields simply say 
\be
	\diff' \CX = 0 \quad  \diff' \CY = 0\,,
\ee	
\ie\, $\CX, \CY$ are constant along $\R$ and holomorphic on $\C$.

Putting this together, $HT$-twisted free hypermultiplets can be succinctly described as an AKSZ theory \cite{AKSZ} based on the following mapping spaces, \cf\, \cite{ESWtax}:
\be
\label{eq:EOMYMHThypers}
\begin{aligned}
	HT^A\textrm{-twisted hypermultiplets: }& \Sect(\R_\dR \times \C_\opd \times \C[-1], T^*\C \otimes K_\C^{1/2})\\
	HT^B\textrm{-twisted hypermultiplets: }& \Maps(\R_\dR \times T^*[1]\C_\opd, T^*[2]\C)\\
\end{aligned}
\ee
In this expression, $M_\dR$ denotes the DG-manifold whose smooth functions are identified with $\Omega^\bullet(M)$ and differential given by the de Rham differential $\diff$ on $M$. Similarly, $Z_\opd$, for $Z$ a complex manifold, denotes the dg-manifold with smooth functions given by $\Omega^{0,\bullet}(Z)$ and differential given by the Dolbeault differential $\opd$ on $Z$. Finally, the notation $\Sect$ is meant to indicate that the equations of motion for the $HT^A$-twisted hypermultiplet is the space of sections of the $\Spin(2)_{E'}$-bundle $T^*\C \otimes K_\C^{1/2}$ over $\R_\dR \times \C_\opd \times \C[-1]$, rather than, say, maps into the total space of the bundle $T^*\C \otimes K_\C^{1/2}$.

We will see that the topological deformations of the $HT$ twist can be realized geometrically. The topological $A$ twist (resp. $B$ twist) is realized by deforming the differential $Q_{HT}$ by $\pd_\varepsilon$ (resp. $-\varepsilon \pd$); this feature is to be expected, see \eg\, \cite{Butson2, BLS, CostelloPSI}. In particular, we find that, at the level of solutions to the equations of motion, the deformations take the following form.
\begin{figure}[H]
	\centering
	\begin{tikzpicture}
		\draw (-4.5,1.5) node {$\Sect(\R_\dR \times \C_\opd \times \C[-1],T^* \C \otimes K_\C^{1/2})$};
		\draw (-4.5,-0.25) node {$\Sect(\R_\dR \times \C_\opd, T^*\C \otimes K^{1/2}_{\C})_{\dR}$};
		\draw[->,decorate,decoration={snake,amplitude=.4mm,segment length=3mm, post length=1mm}] (-3.75,1.25) -- (-4.25,0.25);
		\draw (-6, 0.75) node {deform to $A$ twist};
		
		\draw (4,1.5) node {$\Maps(\R_\dR \times T^*[1]\C_\opd,T^*[2] \C)$};
		\draw (4,-0.25) node {$\Maps(\R^3_\dR, T^*[2]\C)$};
		\draw[->,decorate,decoration={snake,amplitude=.4mm,segment length=3mm, post length=1mm}] (3.75,1.25) -- (4.25,0.25);
		\draw (6, 0.75) node {deform to $B$ twist};
		
		\draw[<->] (-1.25,1.5) -- (1.25,1.5);
		\draw (0,1.75) node {change twist};
		\draw (0,1.25) node {data};
	\end{tikzpicture}
\end{figure}

\subsection{$A$ twist}
\label{sec:FHA}
First consider deforming the above $HT$-twisted theory to the topological $A$ twist. We seek to deform the differential $Q_{HT}$ to obtain a nilpotent transformation $Q_A$ realizing the $A$ twist of free hypermultiplets. Moreover, the total differential $Q_A$ should be realized as taking the BV-bracket with a deformed action. Note that deformations of the action leading to the $A$ twist should have an integrand with $J_A = 1$ and $C_A = 2$, where the twisted spins $J_A$ and cohomological degrees $C_A$ are given in Tables \ref{table:FHspin} and \ref{table:FHcoho}, respectively. For ease of reading, we write them once again in Table \ref{table:FHspincohoA}.

\begin{table}[H]
	\centering
	\begin{tabular}{c|c|c|c|c}
		& $\bX$ & $\bPsi_{\bX}$ & $\bY$ & $\bPsi_{\bY}$ \\ \hline
		$J_A$ & $\tfrac{1}{2}$   & $\tfrac{1}{2}$    & $\tfrac{1}{2}$ & $\tfrac{1}{2}$\\ 
		$C_A$ & $0$   & $1$    & $0$ & $1$\\
	\end{tabular}
	\caption{Twisted spin $J_A$ and cohomological grading $C_A$ of the twisted superfields for free hypermultiplets in the topological $A$ twist.}
	\label{table:FHspincohoA}
\end{table}

A guiding principle in searching for the desired deformation is that it should reflect the action of $-Q^2_- = Q^{-\dot{+}}_-$ on the physical fields. Upon investigating the transformation of the hypermultiplets under $Q^2_-$ found above, one finds that $Q^2_- \psi_X = Q^2_- \psi_Y = 0,$ $Q^2_- X \sim \psi_Y,$ and $Q^2_- Y \sim \psi_X.$ Therefore, the deformation of the action takes the form
\be
\label{eq:AtwistedactionFH}
S_A = S - \int \bPsi_{\bX} \bPsi_{\bY}.
\ee
The resulting action of $Q_A$ is given by
\be
\begin{aligned}
	\label{eq:QAFH}
	Q_A \bX & = \diff' \bX - \bPsi_{\bY} \qquad & Q_A \bPsi_{\bX} & = \diff' \bPsi_{\bX}\\
	Q_A \bY & = \diff' \bY + \bPsi_{\bX} \qquad & Q_A \bPsi_{\bY} & = \diff' \bPsi_{\bY}\\
\end{aligned}\,,
\ee
It is straightforward to check that $S_A$ is $Q_A$ invariant and that the action of $Q_A$ agrees with $\{-,S_A\}_{BV}$. As expected, the deformation to the differential $Q_{HT}$ acts as $\partial_\varepsilon$ on the combined superfields $\CX, \CY$ in Eq. \eqref{eq:combinedhypers}.

Note the above deformation is chiral, thus the results of \cite{GwilliamWilliams} can be freely applied. We find that the (modified) stress tensor superfield $\bT$ for this $A$-twisted theory is given by
\be
	\bT = \tfrac{1}{2}\big(\bPsi_{\bX} \pd_z \bX + \bPsi_{\bY} \pd_z \bY - \bX \pd_z \bPsi_{\bX} - \bY \pd_z \bPsi_{\bY}\big)\,,
\ee
from which it follows that $Q_A \bT = \diff' \bT = \diff \bT$, the second equation holding because $\bT$ has twisted spin $1$, \ie\, it can be identified with a differential form proportional to $\diff z$. Moreover, $\bS = \tfrac{1}{2}(\bY \pd_z \bX - \bX \pd_z \bY)$ satisfies Eq. \eqref{eq:QST} so that our deformed theory is classically topological. The analysis of Section \ref{sec:Qstress} implies that the quantum theory is also topological.

\subsubsection{Costello-Gaiotto boundary algebra}
\label{sec:bdyFHA}
Our first goal will be to use the above twisted theory to recover the boundary algebras of \cite{CostelloGaiotto} for the $A$ twist of a free hypermultiplet, \ie\, the symplectic bosons VOA $\Sb$. This VOA plays a role analogous to the WZW model in the Chern-Simons/WZW correspondence \cite{WittenJones}. For example, the category of line operators is expected to be identified with a certain category of modules for the boundary algebra \cite{CostelloGaiotto, CostelloCreutzigGaiotto, BNbraid}. Moreover, the state space $\CH_A(\Sigma)$, possibly punctured by line operators, should be reproduced by the space of conformal blocks of $\Sb$ on $\Sigma$, with the corresponding modules at the punctures \cite{GaiottoTwisted}; we will return to these state spaces in Section \ref{sec:hilbFHA}.

In terms of twisted superfields, the boundary conditions of interest are given by Neumann boundary conditions for both of the chiral multiplets ($\bPsi_{\bX}|_{\pd}, \bPsi_{\bY}|_{\pd} = 0$), \cf\, \cite[Section 2.4]{CostelloGaiotto}. The only fields that survive $Q_A$ cohomology at the boundary are the lowest components $X,Y$ of the bosonic twisted superfields $\bX, \bY$. Since these fields are {\em not} $Q_A$ closed in the bulk, it follows that they can have a non-trivial OPE at the boundary, \cf\, \cite{CostelloDimofteGaiotto-boundary}. 

The boundary OPE is highly constrained by charge conservation and can be computed in the same fashion as in \cite[Sec 5.3]{CostelloDimofteGaiotto-boundary}. We have propagators represented by the oriented edges
\begin{figure}[H]
	\centering
	\begin{tikzpicture}
		\draw (5+1.9,0) node {$\bX$};
		\draw (5-1.9,0) node {$\bPsi_{\bX}$};
		\draw[middlearrow={latex}, thick] (5-1.41,0) -- (5+1.41,0);
	\end{tikzpicture}
	\qquad
	\begin{tikzpicture}
		\draw (1.9,0) node {$\bPsi_{\bY}$};
		\draw (-1.9,0) node {$\bY$};
		\draw[middlearrow={latex}, thick] (-1.41,0) -- (1.41,0);
	\end{tikzpicture}
	
\end{figure}
\noindent and a single interaction vertex of the form
\begin{figure}[H]
	\centering
	\begin{tikzpicture}
		\draw (5,0) node {$\otimes$};
		\draw (5,0.5) node {$-\int \bPsi_{\bX}\bPsi_{\bY}$};
		\draw[middlearrow={latex}, thick] (5-1.41,0) -- (5,0);
		\draw[middlearrow={latex}, thick] (5,0) -- (5+1.41,0);
	\end{tikzpicture}
\end{figure}
\noindent The highly restrictive structure of these interaction vertices and propagators indicates that this theory is 1-loop exact when using a holomorphic gauge fixing, \cf\, \cite{CostelloDimofteGaiotto-boundary, GwilliamWilliams}. This interaction is ``chiral," using the language of \cite{GwilliamWilliams}. 

Concretely, we must fix the gauge for the ``exotic" symmetry with ghosts $\psi_X, \psi_Y$. This is done as in \cite{CostelloDimofteGaiotto-boundary, GwilliamWilliams} by imposing the gauge fixing condition%
\be
\diff'* \eta_X = \diff'* \eta_Y = 0\,.
\ee
With this choice of gauge, the 2-point function $\bPsi_{\bX}(z,t) \bX(w,s)$ in the absence of the above boundary condition is given by
\be
	P(z,t;w,s) = \frac{(t-s) \diff(\oz - \ow) - (\oz - \ow)\diff(t-s)}{8 \pi i(|z-w|^2 + (t-s)^2)^{3/2}} \diff z^{1/2} \diff w^{1/2}\,.
\ee
and similarly for the 2-point function $\bPsi_{\bY}(z,t) \bY(w,s)$. The Neumann $(\CN)$ boundary condition modifies the 2-point function to
\be
P_\CN(z,t;w,s) = \frac{1}{2}\big(P(z,t;w,s) - P(z,-t;w,s)\big)\,,
\ee
which obviously vanishes if $\bPsi_{\bX}$ or $\bPsi_{\bY}$ is at the boundary $t=0$.

There is a single Feynman diagram that contributes to boundary OPE of $X,Y$:
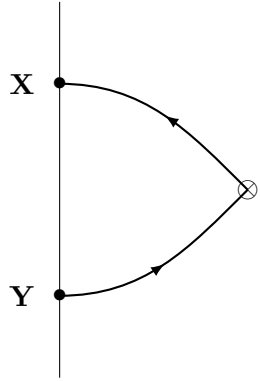
\begin{figure}[H]
	\centering
	\begin{tikzpicture}
		\node (int) at (0,0) {$\otimes$};
		\node (Y) at (-3,-1.41) {$\bY$};
		\node (bY) at (-2.5,-1.41) {$\bullet$};
		\node (X) at (-3,1.41) {$\bX$};
		\node (bX) at (-2.5,1.41) {$\bullet$};
		\draw[middlearrow={latex}, thick] (bY.center) to [in=225, out=0] (int.center);
		\draw[middlearrow={latex}, thick] (int.center) to [in=0, out=135] (bX.center);
		\draw (-2.5,2.5) -- (-2.5,-2.5);
	\end{tikzpicture}
\end{figure}
\noindent Since both $\bX$ and $\bY$ have twisted spin $\tfrac{1}{2}$, this diagram contributes at order $\tfrac{1}{z-w}$. Indeed, we can compute it exactly:
\be
\begin{aligned}
	\int_{x,\ox} \int_{s\geq 0}& P_\CN(x,s;z,0) P_\CN(x,s;w,0)\\
	& = -\frac{\diff z^{1/2} \diff w^{1/2}}{64\pi^2}\int_{x,\ox}  \int_{s\geq 0} \frac{s(\oz-\ow)}{(|x-z|^2 +  s^2)^{3/2}(|x-w|^2 +  s^2)^{3/2}} \diff s \diff \ox \diff x\\
	& = \frac{1}{\pi i}\frac{\diff z^{1/2} \diff w^{1/2}}{z-w}
\end{aligned}
\ee
The fact that the contribution of this diagram is proportional to $\diff z^{1/2} \diff w^{1/2}$ corresponds to the fact that the only non-trivial OPE is between the lowest components $X(z)$ and $Y(w)$. Thus, up to a numerical prefactor, we see that the OPE of $X$ and $Y$ is given by
\be
\label{eq:opeFHA}
X(z) Y(w) \sim \frac{1}{z-w}\,,
\ee
which is simply a copy of the symplectic bosons VOA $\Sb$. As expected, we have reproduced the results of \cite{CostelloGaiotto}.

\subsubsection{State spaces, local operators, and line operators}
\label{sec:hilbFHA}
From the twisted description given above, it is relatively straightforward to extract many interesting aspects of the twisted theory by (shifted) geometric quantization of the spaces of solutions to the equations of motion, \cf\, \cite{EllYooLanglands}. For example, by considering the ($0$-shifted symplectic) space of solutions to the equations of motion on a surface $\Sigma$, geometric quantization yields the quantum state space $\CH(\Sigma)$. These states are expected to be identified with conformal blocks of the symplectic boson VOA described in the previous section \cite{GaiottoTwisted}. In a similar fashion, we can extract the category of line operators by quantizing the ($1$-shifted symplectic) space of solutions to the equations of motion the algebraic avatar of a circle $S^1$, the formal punctured disk $\D^\times$.

We start with $HT^A$ twisted theory. The combined fields $\CX, \CY$, as in Eq. \eqref{eq:combinedhypers}, are each sections of $K_\Sigma^{1/2}$ over $\R \times \Sigma \times \C^{0|1}$ whose equations of motion say that $\diff' \CX = \diff' \CY = 0$. We can therefore identify the space of solutions on $\R \times \Sigma$ with
\be
	\Sect(\Sigma_\opd \times \C[-1], T^*\C \otimes K^{1/2}_{\Sigma}) \cong T[1]\Sect(\Sigma_\opd, T^*\C \otimes K^{1/2}_{\Sigma})\,.
\ee
A map from (or section of a bundle over) $\C[-1]$ can be viewed as a pair of a point and an odd, holomorphic tangent vector at that point. Roughly, $-\bPsi_{\bY}$ can be thought of as the differential $\delta \bX$, and similarly for $\bPsi_{\bX}$ and $\delta \bY$.%
\footnote{More precisely, $\bX, \bY$ are coordinate functions on $\Sect(\Sigma_\opd, T^*\C \otimes K^{1/2}_{\Sigma})$, and $-\bPsi_{\bY}, \bPsi_{\bX}$ are linear functions on the fiber of the odd tangent bundle, \ie\, 1-forms over $ \Sect(\Sigma_\opd, T^*\C \otimes K^{1/2}_{\Sigma})$.}

The $A$ twist deformation corresponds to including the differential $\pd_\varepsilon$, geometrically corresponding to including the (holomorphic) de Rham differential to the $HT$-supercharge. This deforms the space of solutions to the equations of motion as
\be
\label{eq:EOMFHA}
	T[1]\Sect(\Sigma_\opd, T^*\C \otimes K^{1/2}_{\Sigma}) \rightsquigarrow \Sect(\Sigma_\opd, T^*\C \otimes K^{1/2}_{\Sigma})_\dR \cong T^* \Sect(\Sigma_\opd, K^{1/2}_{\Sigma})_\dR.
\ee
The state space of the $A$ twist on $\Sigma$ then corresponds to geometric quantization of this ($0$-shifted) cotangent bundle. Working with the natural polarization induced by the fact this is a cotangent bundle, find that geometric quantization results in functions on the base $\Sect(\Sigma_\opd, K^{1/2}_{\Sigma})_\dR$, which is simply (Borel-Moore)%
\footnote{The homology theory one should work with is quite subtle, especially once we include gauge fields in Section \ref{sec:SYM}. As described in \cite[Sec 6]{SafronovWilliams} for general $A$-twisted $\sigma$-models, and in particular Example 6.5 of \emph{loc. cit.}, the correct choice for the present case is Borel-Moore homology. Indeed, this homology theory was originally proposed by Nakajima \cite{Nak} and later used by Braverman-Finkelberg-Nakajima \cite{BFNII} to realize a mathematically precise definition of Coulomb branches of the 3d $\CN=4$ theories described in this paper. In contrast to more traditional homology, Borel-Moore homology allows cycles to be non-compact and is dual to compactly-supported cohomology. See \eg\, \cite[Ch 2]{ChrissGinzburg} for details about Borel-Moore homology in the context of geometric representation theory.} %
homology of $\Sect(\Sigma_\opd,K^{1/2}_{\Sigma})$:
\be
	\CH_A(\Sigma) = H_\bullet\big(\Sect(\Sigma_\opd, K^{1/2}_{\Sigma})\big)\,,
\ee
\cf\, \cite{GaiottoTwisted, BFKHilb, SafronovWilliams}.

We can easily determine the algebra of local operators in the $A$ twist in two different ways. First, we could consider a fully non-perturbative analysis via a state-operator correspondence, \ie\, identifying local operators with states on $S^2$, or better a raviolo $\D \cup_{\D^\times} \D$. Alternatively, we can consider the purely perturbative analysis of building $Q_A$-cohomology classes out of the fundamentals fields. In both cases we reproduce the known result of a trivial algebra, \ie\, only the identity operator $\mathds{1}$.

The non-perturbative analysis goes as follows. A point in $\Sect(\Sigma_\opd, K^{1/2}_{\Sigma})$ for $\Sigma = \D \cup_{\D^\times} \D$ is identified with a pairs of holomorphic spinors $(X(z), X'(z))$ over $\D$ that agree over $\D^\times$. The spinor $X(z)$, and similarly $X'(z)$, can be identified with a formal Taylor series $X(z) \in \C[\![z]\!]$, with the understanding that the series transforms as $X(z) \to \lambda^{1/2} X(\lambda z)$ under the (complexified) spacetime rotation $z \to \lambda z$.%
\footnote{We could alternatively choose a slightly different twisting homomorphism where $\bX$ is a scalar and $\bY$ is a section of $K_\Sigma$. This corresponds to working with the twisted spin $\wt{J}_A = J_A - \tfrac{1}{2} M$.} %
Since this space of sections is contractible, we conclude that the space of local operators is simply the Borel-Moore homology of a point, \ie\,
\be
\label{eq:opsFHA}
	\CH_A(\D \cup_{\D^\times} \D) = H_\bullet\big({\rm point}\big) = \C\,.
\ee

We now consider the perturbative analysis of building bulk local operators out of products of the $Q_A$-closed twisted superfields, up to $Q_A$-exact terms. We can determine the structure of this cohomology by a spectral sequence. First, we consider the cohomology of $Q_{HT}$. This is described in \cite{OhYagi, CostelloDimofteGaiotto-boundary} and can be readily seen to be products of the lowest components $X, \psi_X, Y, \psi_Y$ and their $z$-derivatives.

The differential $\delta_A$ on the next page of the spectral sequence is the remainder of $Q_A$
\be
\begin{aligned}
	\label{eq:pg2FHA}
	\delta_A X & = -\psi_Y \qquad & \delta_A \psi_{X} & = 0\\
	\delta_A Y & = \psi_X \qquad & \delta_A \psi_{Y} & = 0\\
\end{aligned}\,.
\ee
We are forced to remove the pairs $(X, -\psi_Y)$ and $(Y, \psi_X)$ of exact terms and their primitives, and the spectral sequence terminates. Thus, the cohomology is simply generated by the identity operator $\mathds{1}$, as expected.

The above analysis admits a straightforward generalization to the inclusion of line operators. In general, the category of line operators can be obtained by geometric quantization of the ($0-$shifted symplectic) space of solutions to the equations on motion on the $S^1$ link of the line operator. The algebraic avatar of $S^1$ is the (formal) punctured disk $\D^\times$, whence we consider the above analysis with $\Sigma = \D^\times$. Namely, start with the equations of motion on the formal disk $\D^\times$, \ie\, $\Sect(\D^\times_\opd, T^*\C \otimes K^{1/2}_{\D^\times})_\dR \cong T^*[1]\Sect(\D^\times_\opd, K^{1/2}_{\D^\times})_\dR$. Again, the base of this ($1$-shifted) cotangent bundle can be used as a polarization. It is also convenient to use the above trivialization of $K^{1/2}$, \ie\, identify $\Sect(\D^\times_\opd, K^{1/2}_{\D^\times})$ with functions on the (algebraic) loop space $\CL \C$. Geometric quantization then yields \cite{SafronovGQ} quasi-coherent sheaves on the base:
\be
	\CC_A = \textrm{QCoh}(\CL\C_\dR) \cong \textrm{D}(\CL \C)
\ee
where $\CC_A$ denotes the category of line operator (in the $A$ twist) and $\textrm{D}(X)$ denotes the (derived) category of $D$-modules on $X$. Indeed, this is the expected answer for the category of line operators in the $A$-twist of a free hypermultiplet, \cf\, \cite{linevortex, CostelloGaiotto, HilburnRaskin}.

\subsubsection{Descent}
\label{sec:descentFHA}
Given the form of the $A$-twisted action, we can attempt to perform the analysis discussed in Section \ref{sec:descent}. Since the algebra of local operators in the $A$-twist is trivial, this bracket should also be trivial. Nonetheless, the operators $\psi_X, \psi_Y$ are both $Q_A$- and $Q_{HT}$-closed and therefore it makes sense to determine their topological descent brackets in terms of some holomorphic-topological descent brackets.

The variations given in Eq. \eqref{eq:QFH} and Eq. \eqref{eq:QAFH} imply that
\be
(\psi_X^{(1)})^{\langle 1 \rangle} = \pd_z Y^{(1)} \wedge \diff z \qquad (\psi_Y^{(1)})^{\langle 1 \rangle} = -\pd_z X^{(1)} \wedge \diff z.
\ee
From this we get
\be
\begin{aligned}
	\{\!\{\psi_X, \psi_Y\}\!\} = \{\!\{\pd_z Y, \psi_Y\}\!\}_{HT} = 0\\
	\{\!\{\psi_Y, \psi_X\}\!\} = -\{\!\{\pd_z X, \psi_X\}\!\}_{HT} = 0\\
\end{aligned}\,.
\ee
Of course, the vanishing of these brackets is to be expected as $\psi_X, \psi_Y$ are both $Q_A$ exact.

\subsubsection{Deformations induced by flavor symmetries}
\label{sec:FHAdef}
We will end this section by noting the $A$ twist of a free hypermultiplet can be deformed by background fields that couple to the $SU(2)_M$ flavor symmetry. 

We can make the full $SU(2)_M$ flavor symmetry manifest by denoting $\bZ^1 = \bX, \bZ^2 = \bY$ and $\bPsi_1 = \bPsi_{\bX}, \bPsi_2 = \bPsi_{\bY}$. The $A$ twist action then takes the form
\be
S_A = \int \bPsi\diff' \bZ + \tfrac{1}{2} \Omega^{-1}(\bPsi, \bPsi)\,,
\ee
where $\Omega^{-1}$ is Poisson tensor on $T^*\C \cong \C^2$. In this notation, the action of $Q_A$ is expressed as
\be
Q_A \bZ^n = \diff' \bZ^n + \Omega^{mn} \bPsi_m \qquad Q_A \bPsi_m = \diff' \bPsi_m\,.
\ee
We can turn on a background fields in an $\CN=4$ vector multiplet for the $\mathfrak{su}(2)_M$ flavor symmetry by introducing the bosonic twisted superfields $\widehat{\bA} \in \bOmega^{\bullet, (0)} \otimes \mathfrak{su}(2)_M[1]$ and $\widehat{\bPhi} \in~\bOmega^{\bullet, (0)} \otimes \mathfrak{su}(2)_M[2]$ and the deformed action
\be
\label{eq:FHAdefaction}
	S_A \rightsquigarrow S_A + \int \bPsi \widehat{\bA} \bZ + \tfrac{1}{2} \Omega(\bZ, \widehat{\bPhi} \bZ)\,.
\ee
This deformation solves the classical master equation so long as the constraints
\be
	F'(\widehat{\bA}) + \widehat{\bPhi} = 0 \qquad \diff'_{\widehat{\bA}} \widehat{\bPhi} = 0
\ee
are satisfied, agreeing with the fact that the $A$ twist localizes to monopole configurations.%
\footnote{The cohomological degree 0 fields are the 1-form part of $\widehat{\bA}$ and the 2-form part of $\widehat{\bPhi}$. The latter can be identified with the auxiliary field in the adjoint chiral multiplet inside the $\CN=4$ multiplet, which is equal to $\omu_\C$ on-shell. Thus, noting that $(\widehat{\bA})|_{1\textrm{-form}} = (\widehat{A}_t -i \widehat{\sigma}) \diff t + \widehat{A}_\oz \diff \oz$, the first equation reads $F_{\oz t} - i D_\oz \sigma  \sim \omu_\C$, which is exactly the $\diff \oz \diff t$ part of the Bogomolny equations.} %
Perhaps unsurprisingly, we will see in Section \ref{sec:SYMA} that these equations become the equations of motion for $\widehat{\bA}$ and $\widehat{\bPhi}$ in the $A$ twist.

For the special case $\widehat{\bPhi} = 0$, the constraints implies that $\widehat{\bA}$ is a connection on a $SU(2)_M$ bundle that is constant in time and holomorphic on the surface $\Sigma$. It is important to note that turning such on a background holomorphic $SU(2)_M$ bundle $E \to \Sigma$ doesn't modify the OPEs of the boundary symplectic boson VOA. Rather, it modifies what constitutes as a consistent correlation function -- the symplectic bosons no longer transform as spinors on the surface $\Sigma$, \ie\, sections on $K^{1/2}_\Sigma \otimes \C^2$, but instead as spinors twisted by $E$, \ie\, sections on $K^{1/2}_\Sigma \otimes E_{\C^2}$ for $E_{\C^2} = E \times_{SU(2)_M} \C^2$ the associated bundle. Thus, the existence of this deformation by a background holomorphic bundle implies that the state spaces $\CH_A(\Sigma, E)$ define a sheaf over $\Bun_{SL(2,\C)}(\Sigma)$. An even more precise statement is that $\CH_A(\Sigma, E)$ admits the structure of a $D$-module over $\Bun_{SL(2,\C)}(\Sigma)$. This $D$-module structure is most easily seen by identifying $\CH_A(\Sigma, E)$ with the space of conformal blocks on $\Sigma$ for the symplectic boson VOA $\Sb$ twisted by $E$. This VOA has $\mathfrak{su}(2)_{-1/2}$ currents $J^{(mn)} = \norm{Z^m Z^n}$ implementing to the $SU(2)$ flavor symmetry. It is these $\mathfrak{su}(2)_{-1/2}$ currents that realize the $D$-module structure on the sheaf of state spaces, \cf\, \cite{GaiottoTwisted}.

\subsection{$B$ twist}
\label{sec:FHB}
Now consider the deformation of the $HT^B$ twist to the topological $B$ twist. Note that deformations of the action leading to $\delta_B$ should be the integral of an expression with $J_B = 1$ and $C_B = 2$. We collect the twisted spin and cohomological grading of the twisted superfields in Table \ref{table:FHspincohoB}.

\begin{table}[H]
	\centering
	\begin{tabular}{c|c|c|c|c}
		& $\bX$ & $\bPsi_{\bX}$ & $\bY$ & $\bPsi_{\bY}$ \\ \hline
		$J_B$ & $0$   & $1$    & $0$ & $1$\\
		$C_B$ & $0$   & $1$    & $2$ & $-1$ 
	\end{tabular}
	\caption{Twisted spin $J_B$ and cohomological grading $C_B$ of the twisted superfields for free hypermultiplets in the topological $B$ twist.}
	\label{table:FHspincohoB}
\end{table}

Upon investigating the transformation of the hypermultiplets under $\oQ^2_-$, one finds that $\oQ^2_- X = \oQ^2_- Y = 0,$ $\oQ^2_- \psi_X \sim \pd_z Y,$ and $\oQ^2_- \psi_Y \sim \pd_z X.$ We expect that the deformation of the action should be of the form
\be
\label{eq:BtwistedactionFH}
S_B = S + \int\bY \pd \bX.
\ee
The action of the supercharge $Q_B$ is then given by
\be
\begin{aligned}
	\label{eq:QBFH}
	Q_B \bX & = \diff' \bX \qquad & Q_B \bPsi_{\bX} & = \diff' \bPsi_{\bX} - \pd \bY\\
	Q_B \bY & = \diff' \bY \qquad & Q_B \bPsi_{\bY} & = \diff' \bPsi_{\bY} + \pd \bX\\
\end{aligned}\,.
\ee
It is straightforward to check that $S_B$ is $Q_B$ invariant and that the action of $Q_B$ agrees with $\{-,S_B\}_{BV}$. We note that, in terms of the combined fields $\CX, \CY$, the deformation is simply $- \varepsilon \partial$. Moreover, the $B$-twisted action takes the following simple form:
\be
\label{eq:BtwistedactionFH2}
	S_B = \int \CY \diff \CX\,,
\ee
exactly reproducing the AKSZ theory based on the mapping space $\Maps(\R^3_\dR, T^*[2]\C)$, \cf\, \cite{KQZ}.

Again, the above deformation is chiral so we can apply the results of \cite{GwilliamWilliams}. The (modified) stress tensor superfield $\bT$ in the $B$-twist is given by
\be
	\bT = \bPsi_{\bX} \pd_z \bX + \bPsi_{\bY} \pd_z \bY,
\ee
from which it follows that $Q_B \bT = \diff \bT$ since $\bT$ has twisted spin $1$. Moreover, we find that $\bS = -\bPsi_{\bX} (\iota_{\pd_z} \bPsi_{\bY})$ satisfies Eq. \eqref{eq:QST} so this theory is classically topological. Just as with the $A$ twist described earlier, the topological nature of the theory survives perturbation theory.

\subsubsection{Costello-Gaiotto boundary algebra}
\label{sec:bdyFHB}
We now consider the boundary algebra of \cite{CostelloGaiotto} for $B$-twisted hypermultiplets. In terms of 3d $\CN = 2$ fields, the boundary conditions of interest are given by Dirichlet boundary conditions for both of the chiral multiplets ($\bX|_{\pd}, \bY|_{\pd} = 0$), \cf\, \cite[Section 2.4]{CostelloGaiotto}.

The perturbative analysis of the boundary algebra is nearly identical to the $A$ twist discussed in Section \ref{sec:bdyFHA}. In the absence of the Dirichlet boundary condition, the $\bPsi_{\bX}(z,t) \bX(w,s)$ propagator takes the form
\be
	P(z,t;w,s) = \frac{(t-s) \diff(\oz - \ow) - (\oz - \ow)\diff(t-s)}{8\pi i(|z-w|^2 + (t-s)^2)^{3/2}} \diff z\,,
\ee
and similarly for $\bPsi_{\bY}(z,t) \bY(w,s)$. The propagator in the presence of a Dirichlet $(\CD)$ boundary condition is modified to
\be
P_\CD(z,t;w,s) = \frac{1}{2}\big(P(z,t;w,s) - P(z,t;w,-s)\big)\,,
\ee
which obviously vanishes if $\bX$ or $\bY$ is at the boundary $s=0$.

The superpotential $\bY \pd \bX$ induces single interaction vertex. Just as with the $A$ twist, this is a ``chiral" interaction and the Feynman diagram analysis in holomorphic gauge is 1-loop exact. The only boundary local operators that survive cohomology are the lowest components $\psi_X, \psi_Y$ of the fermionic fields $\bPsi_{\bX}, \bPsi_{\bY}$. Again, as they are not $Q_B$ closed in the bulk, they can have a non-trivial OPE on the boundary computed from the following diagram.
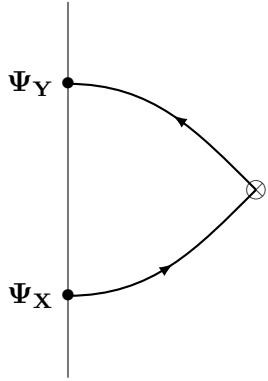
\begin{figure}[H]
	\centering
	\begin{figure}[H]
		\centering
		\begin{tikzpicture}
			\node (int) at (0,0) {$\otimes$};
			\node (psiX) at (-3,-1.41) {$\bPsi_{\bX}$};
			\node (bpsiX) at (-2.5,-1.41) {$\bullet$};
			\node (psiY) at (-3,1.41) {$\bPsi_{\bY}$};
			\node (bpsiY) at (-2.5,1.41) {$\bullet$};
			\draw[middlearrow={latex}, thick] (bpsiX.center) to [in=225, out=0] (int.center);
			\draw[middlearrow={latex}, thick] (int.center) to [in=0, out=135] (bpsiY.center);
			\draw (-2.5,2.5) -- (-2.5,-2.5);
		\end{tikzpicture}
	\end{figure}
\end{figure}
\noindent
This leads to the OPE (up to a non-zero numerical factor)
\be
\label{eq:opeFHB}
\psi_X(z) \psi_Y(w) \sim \frac{1}{(z-w)^2},
\ee
exactly reproducing the OPE for the fermionic currents VOA $\Fc$, equivalently the affine $\rm{psu}(1|1)_1$ VOA.

\subsubsection{State spaces, local operators, and line operators}
\label{sec:hilbFHB}
We now turn to the state spaces of a $B$-twisted hypermultiplet on a Riemann surface $\Sigma$ using the $B$-twisted description above. Just as the $A$-twisted state spaces discussed in Section \ref{sec:hilbFHA} reproduce spaces of conformal blocks for the symplectic boson VOA, these $B$-twisted state spaces should reproduce the spaces of conformal blocks for the affine $\rm{psu}(1|1)_1$ VOA. We proceed by geometric quantization of the solutions to the equations of motion on $\R \times \Sigma$, \cf\, \cite{EllYooLanglands}.

The space of solutions to the equations of motion in the $HT^B$ twist can be identified with the following space of maps
\be
\Maps(T^*[1]\Sigma_\opd, T^*[2] \C) \cong \Maps(\Sigma_\Dol, T^*[2] \C)\,.
\ee
where $\Sigma_\Dol$ is the DG-manifold whose space of smooth functions is the full Dolbeault complex $\Omega^{\bullet, \bullet}(\Sigma)$. As mentioned above, the $B$ twist deformation corresponds to including the additional differential $-\varepsilon \pd$. From this perspective, the deformation to the $B$ twist simply corresponds to deforming $\Sigma_\Dol$ to $\Sigma_\dR$, thus the space of solutions to the equations of motion deforms as
\be
	\Maps(\Sigma_\Dol, T^*[2] \C) \rightsquigarrow \Maps(\Sigma_\dR, T^*[2] \C)\,.
\ee

Geometric quantization of this ($0$-shifted) cotangent bundle is again fairly straightforward, although we have some flexibility in choice of polarization. We note that the complex structure on $\Sigma$ induces the following exact sequence at the level of the ring of functions:
\be
	0 \to (\Omega^{\bullet,1}(\Sigma), \opd) \hookrightarrow (\Omega^{\bullet, \bullet}(\Sigma), \pd + \opd) \twoheadrightarrow (\Omega^{\bullet, 0}(\Sigma), \opd) \to 0
\ee
The projection map corresponds to the inclusion $\Maps(\Sigma_\opd, T^*[2]\C) \hookrightarrow \Maps(\Sigma_\dR, T^*[2]\C)$, which is Lagrangian. Geometric quantization then leads to a state space given by functions on the mapping space $\Maps(\Sigma_\opd, T^*[2]\C)$, which can be identified with functions on $H^{\bullet,0}(\Sigma) \otimes T^*[2]\C \cong (\C \oplus \C[1]^{g})\otimes T^*[2]\C$. The even generators arrange themselves into functions on $T^*[2]\C$ and odd generators can be viewed as sections of the (shifted) holomorphic tangent bundle $T[1] (T^*[2]\C)$. We thus conclude the state space is given by 
\be
\label{eq:hilbFHB}
	\CH_B(\Sigma) \cong H^\bullet_\opd\big(T^*[2]\C, \wedge^\bullet T[1](T^*[2]\C)^{\otimes g})\big)\,,
\ee
where $g$ is the genus of $\Sigma$, as expected for Rozansky-Witten theory with target $T^*\C$ \cite{RW, BFKHilb}. This general result implies that the vector space of local operators, \ie\, $\CH_B(S^2)$, is
\be
	\CH_B(S^2) \cong H^\bullet_\opd(T^*[2]\C) \cong \C[X,Y]\,.
\ee 

We now turn to a perturbative analysis of the algebra of local operators. This can be again determined by a spectral sequence. First, we consider the cohomology of $Q_{HT}$, which is again generated by the bottom components $X, \psi_X, Y, \psi_Y$ and their $z$-derivatives. The differential on the next page of the spectral sequence is
\be
\begin{aligned}
	\label{eq:pg2FHB}
	\delta_B X & = 0 \qquad & \delta_B \psi_{X} & = -\pd_z Y\\
	\delta_B Y & = 0 \qquad & \delta_B \psi_{Y} & = \pd_z X\\
\end{aligned}\,.
\ee
Operators built from the fields $X, Y$ are closed but not exact. and we can remove the pairs $(\psi_X, \pd_z Y)$ and $(\psi_Y, \pd_z X)$ of exact terms and their primitives. Moreover, the spectral sequence terminates at this page so that we conclude the algebra of local operators is simply $\C[X,Y]$.

We can also realize the category of line operators from geometric quantization as we did in the $A$-twist. We start with the equations of motion on the (formal) punctured disk $\Sigma = \D^\times$:
\be
	\Maps(\D^\times_{\dR}, T^*[2]\C) \cong T^*[1] \Maps(\D^\times_\dR, \C)
\ee
To arrive at the category of lines in \cite{linevortex, HilburnRaskin}, we consider the polarization of this shifted cotangent bundle induced by the base. Geometric quantization yields quasi-coherent sheaves on this space of maps:
\be
	\CC_B = \textrm{QCoh}(\Maps(\D^\times_\dR, \C))
\ee

\subsubsection{Descent}
\label{sec:descentFHB}
Given the form of the $B$-twisted action, we can easily perform the analysis discussed in Section \ref{sec:descent}. The zero-modes of $X, Y$ are $Q_B$- and $Q_{HT}$-closed, so we can hope to find their topological descent brackets in terms of some holomorphic-topological descent brackets. We find that variations given in Eq. \eqref{eq:QFH} and Eq. \eqref{eq:QBFH} imply that
\be
(X^{(1)})^{\langle 1 \rangle} = \psi_Y^{(1)} \wedge \diff z \qquad (Y^{(1)})^{\langle 1 \rangle} = -\psi_X^{(1)} \wedge \diff z
\ee
which imply that 
\be
\begin{aligned}
	\{\!\{X, Y\}\!\} = \{\!\{\psi_Y, Y\}\!\}_{HT} = 1 \\
	\{\!\{Y, X\}\!\} = -\{\!\{\psi_X, X\}\!\}_{HT} = -1\\
\end{aligned}.
\ee
This is the expected result for the $B$ twist of a free hypermultiplet: the topological descent bracket reproduces the Poisson bracket on $T^*[2]\C$ \cite{Yagi, descent}.

\subsubsection{Deformations induced by flavor symmetries}
\label{sec:FHBdef}
We can once again turn on background fields for the $SU(2)_M$ flavor symmetry of our $B$-twisted hypermultiplet. The $B$ twist of an $\CN=4$ vector multiplet localizes to complexified flat connections, thus we expect the $B$ twist of a free hypermultiplet to admit deformations by background flat $SL(2,\C)_M$ connections. The appearance of flat $SL(2,\C)$ connections in a $B$-twisted free hypermultiplet is well known, and the partition function on a 3-manifold $M_3$ with generic flat $SL(2,\C)_M$ connection $\CA$ reproduces the Reidemeister-Ray-Singer torsion of $\CA$ \cite{Mikhaylov, RStorsion}. 

Being able to deform a 3d TQFT by flat $G_\C$ connections is particularly special; in essence, observables fiber over the space of flat connections. A choice of background flat connections for abelian flavor symmetries, and the related $\Spin^c$ structures, play a central role in the partition function analysis of \cite{GHNPPS}. Similarly, the state space of a $B$-twisted 3d $\CN=4$ theory on $\Sigma$ in the presence of the flat connection $\CA$ for the (Higgs branch) flavor symmetry yields a coherent sheaf over the moduli space $\Loc_{G_\C}(\Sigma)$ of flat $G_\C$ bundles (or local systems) on $\Sigma$ \cite{GaiottoTwisted}. Finally, the category of line operators $\CC$ factors into blocks $\CC_\CA$ corresponding to a choice of background flat connection $\CA$; the work \cite{CDGG} relates these blocks to the notion of a relative modular category \cite{DRnonsemisimple} and in recent mathematical works on 3d TQFTs based on quantum groups, \eg\, \cite{BGPMRholonomy, BCGPM, DRGPM}.

A particularly clean way to see the deformation by a flat $SL(2,\C)$ connection $\CA$ is to express the fields in terms of $\CX^1 = \CX$ and $\CX^2 = \CY$. The $B$-twisted action then takes the form
\be
S_B = \tfrac{1}{2}\int\Omega(\CX, \diff \CX)
\ee
from which it is obvious how to include the flat connection $\widehat{\CA}$: we simply replace $\diff$ with the covariant exterior derivative $\diff_\CA = \diff + \widehat{\CA}$.

This has a natural description in terms of $\bX^n, \bPsi_m$; we decompose the flat connection $\widehat{\CA}$ into $\widehat{\bA} \in \bOmega^{1, (0)} \otimes \mathfrak{g}[1]$ and $\widehat{\bPhi} \in \bOmega^{0, (1)} \otimes \mathfrak{g}$. The deformation to the action takes the same form as above:
\be
\label{eq:FHBdefaction}
S_B \rightsquigarrow S_B + \int \bPsi \widehat{\bA} \bX + \tfrac{1}{2}\Omega(\bX,\widehat{\bPhi} \bX)\,,
\ee
\cf\, Eq. \eqref{eq:FHAdefaction}. This action is consistent, \ie\, solves the classical master equation, so long as $\widehat{\bA}, \widehat{\bPhi}$ satisfy
\be
F'(\widehat{\bA}) = 0 \qquad \diff_{\widehat{\bA}} \widehat{\bPhi} - \pd \widehat{\bA} = 0\,,
\ee
which is simply a rewriting that $\CA$ is a flat connection. Again, these constraints will become the equations of motion of $\widehat{\bA}, \widehat{\bPhi}$ in the $B$ twist.

If we work in a holomorphic gauge, \ie\, a gauge where $\widehat{\bA} = 0$ and $\bPhi = \widehat{A}(z) \diff z$, the effect of this deformation becomes particularly transparent. For example, the algebra of local operators bound to a Dirichlet boundary condition deforms as
\be
\psi_n(z) \psi_n(w) \sim \frac{\Omega_{nm}}{(z-w)^2} \quad \rightsquigarrow \quad \psi_n(z) \psi_m(w) \sim \frac{\Omega_{nm}}{(z-w)^2} + \frac{\widehat{A}_{nm}(w)}{z-w}\,,
\ee
where $\widehat{A}_{nm} = \Omega_{nl} \widehat{A}^l{}_m = \widehat{A}_{mn}$, \cf\, \cite{GaiottoTwisted}. As noted in \cite{CDGG}, the deformed VOA for sufficiently generic $\widehat{A}_{MN}$ is simply a copy of free fermions. Free fermions have a trivial category of modules, thus one expects that $\CC_{\widehat{\CA}}$ is trivial for generic $\widehat{\CA}$, in agreement with results of \cite{GHNPPS, Mikhaylov}.

\section{Yang-Mills Gauge Theories}
\label{sec:SYM}
In this section we consider the more sophisticated example of $\CN=4$ hypermultiplets gauged with $\CN=4$ vector multiplets. We assume that the hypermultiplets transform in the representation $T^* V := V \oplus V^*$, for $V$ a unitary representation of the (complexified) gauge group $G$. The deformations to the $A$ and $B$ twists are described in Section \ref{sec:SYMA} and Section \ref{sec:SYMB}, respectively.

As described in Section \ref{sec:N=4asN=2}, the $HT$-twisted theory has twisted superfields $(\bA, \bB), (\bPhi, \bLambda)$ coming from an $\CN=4$ vector multiplet, and $(\bX, \bPsi_{\bX}), (\bY, \bPsi_{\bY})$ coming from the $\CN=4$ hypermultiplets. In addition, the theory has a superpotential of the form $W = -Y \phi X$. The $HT$-twisted action is then given by
\be
\label{eq:twistedactionSYM}
S = \int \bB F'(\bA) + \bLambda \diff'_{\bA} \bPhi+ \bPsi_{\bX} \diff'_{\bA} \bX + \bPsi_{\bY} \diff'_{\bA} \bY - \bY \bPhi \bX,
\ee
and the action of $Q_{HT}$ is given by 
\be
\begin{aligned}
	\label{eq:QSYM}
	Q_{HT} \bA & = F'(\bA) \qquad & Q_{HT} \bB & = \diff'_{\bA} \bB - \bmu\\
	Q_{HT} \bPhi & = \diff'_{\bA} \bPhi & Q_{HT} \bLambda & = \diff'_{\bA} \bLambda - \bmu_\C\\
	Q_{HT} \bX & = \diff'_{\bA} \bX \qquad & Q_{HT} \bPsi_{\bX} & = \diff'_{\bA} \bPsi_{\bX} - \bY \bPhi\\
	Q_{HT} \bY & = \diff'_{\bA} \bY \qquad & Q_{HT} \bPsi_{\bY} & = \diff'_{\bA} \bPsi_{\bY} - \bPhi \bX\\
\end{aligned}\,,
\ee
where $\bmu_\C = \mu_\C(\bX, \bY)$ is the moment map for the $\mathfrak{g}$ action on $T^*V$. In components, the moment maps read%
\be
\bmu_a = \bLambda_c f^c{}_{ab} \bPhi^b + \bPsi_{\bX}{}_m (\tau_a)^m{}_n \bX^n - \bY_m (\tau_a)^m{}_n \bPsi_{\bY}^n \qquad (\bmu_\C)_a = \bY_m (\tau_a)^m{}_n \bX^n\,.
\ee
We will denote the lowest components of the hypermultiplet twisted superfields as in Section \ref{sec:hypers} and the lowest components of the vector multiplet twisted superfields $(\bA,\bB)$ (resp. $(\bPhi, \bLambda)$) by $(c, B)$ (resp. $(\phi, \lambda)$). The brackets arising from holomorphic-topological descent are given by those in Eq. \eqref{eq:descentFH} and%
\footnote{Strictly speaking, these are the brackets in the perturbative algebra of local operators in the $HT$-twist of a free $\CN=4$ vector multiplet. Nonetheless, we expect that these brackets can be used in the perturbative algebra of local operators in the $HT$-twisted interacting theory described above, \cf\, \cite[Sec 3.4]{CostelloDimofteGaiotto-boundary}.} %
\be
\label{eq:descentPG}
\{\!\{c,B\}\!\}_{HT} = \{\!\{\phi,\lambda\}\!\}_{HT} = 1.
\ee

Just as with the $\CN=4$ hypermultiplets, we can combine the twisted superfields coming from the $\CN=4$ vector multiplet as
\be
\label{eq:combinedvect}
\begin{aligned}
	\CA & = \bA + \varepsilon \bPhi \qquad & \CB & = \bLambda + \varepsilon \bB\\
\end{aligned},
\ee
where $\varepsilon$ is as above. We can perform the same analysis as in Section \ref{sec:hypers} to identify the $HT^A$ and $HT^B$ twist of super Yang-Mills with AKSZ theories:
\be
\label{eq:EOMYMHTSYM}
\begin{aligned}
	HT^A\textrm{-twist: }& \Sect(\R_\dR \times \C_\opd \times \C[-1], T^*(V/G) \otimes K_\C^{1/2})\\
	HT^B\textrm{-twist: }& \Maps(\R_\dR \times T^*[1]\C_\opd, T^*[2](V/G))\\
\end{aligned}
\ee
We will again find that the deformation of the $HT$ twist to the topological $A$ twist (resp. $B$ twist) is reproduced by deforming the differential on this mapping space by $\pd_\varepsilon$ (resp. $- \varepsilon \pd$). In terms of mapping spaces, the deformations take the following form.

\begin{figure}[H]
	\centering
	\begin{tikzpicture}
		\draw (-4.75,1.5) node {$\Sect(\R_\dR \times \C_\opd \times \C[-1],T^* (V/G) \otimes K_\C^{1/2})$};
		\draw (-4.5,-0.25) node {$\Sect(\R_\dR \times \C_\opd, T^*(V/G) \otimes K^{1/2}_{\C})_{\dR}$};
		\draw[->,decorate,decoration={snake,amplitude=.4mm,segment length=3mm, post length=1mm}] (-3.75,1.25) -- (-4.25,0.25);
		\draw (-6, 0.75) node {deform to $A$ twist};
		
		\draw (4.15,1.5) node {$\Maps(\R_\dR \times T^*[1]\C_\opd,T^*[2](V/G))$};
		\draw (4.15,-0.25) node {$\Maps(\R^3_\dR, T^*[2](V/G))$};
		\draw[->,decorate,decoration={snake,amplitude=.4mm,segment length=3mm, post length=1mm}] (3.75,1.25) -- (4.25,0.25);
		\draw (6, 0.75) node {deform to $B$ twist};
		
		\draw[<->] (-1.1,1.5) -- (1.1,1.5);
		\draw (0,1.75) node {change twist};
		\draw (0,1.25) node {data};
	\end{tikzpicture}
\end{figure}

\subsection{$A$ twist}
\label{sec:SYMA}
First consider deforming the $HT$ twist of super Yang-Mills to the topological $A$ twist. The twisting data takes the same form as in Section \ref{sec:FHA}, \ie\, we redefine the cohomological grading and twisted spin as in Eq. \eqref{eq:Amixing}. The resulting twisted spin and cohomological grading of our fields are given below. 

\begin{table}[h!]
	\centering
	\begin{tabular}{c|c|c|c|c|c|c|c|c}
		& $\bA$   & $\bB$  & $\bPhi$ & $\bLambda$ & $\bX$ & $\bPsi_{\bX}$ & $\bY$ & $\bPsi_{\bY}$ \\ \hline
		$J_A$ & $0$ & $1$ & $0$ & $1$ & $\tfrac{1}{2}$   & $\tfrac{1}{2}$    & $\tfrac{1}{2}$ & $\tfrac{1}{2}$\\
		$C_A$ & $1$ & $0$ & $2$ & $-1$ & $0$   & $1$    & $0$ & $1$\\
	\end{tabular}
	\caption{Twisted spin $J_A$ and cohomological degree $C_A$ of the twisted superfields in the topological $A$ twist of super Yang-Mills.}
	\label{table:Aspincoho}
\end{table}

Just as with hypermultiplets, we can identify the deformation of the above $HT$-twisted theory to an $A$-twisted theory by investigation of the transformations in the physical theory. Upon investigating the vector multiplet one finds that, $Q^2_- (F_{zt} + i D_z \sigma) 
\sim Q^2_- \varphi = 0,$ $Q^2_- (A_t - i \sigma) \sim \lambda_+$ and $Q^2_- \lambda_- \sim F_{zt} + i D_z \sigma$. The action of $Q^2_-$ on the hypermultiplet is the same as in Section \ref{sec:FHA}. Therefore, we conclude that the deformed action is given by
\be
\label{eq:AtwistedactionSYM}
S_A = S + \int\bB \bPhi - \bPsi_{\bX} \bPsi_{\bY}.
\ee
The resulting action of $Q_A$ is given by
\be
\begin{aligned}
	\label{eq:QASYM}
	Q_A \bA & = F'(\bA) + \bPhi \qquad & Q_A \bB & = \diff'_{\bA} \bB - \bmu\\
	Q_A \bPhi & = \diff'_{\bA} \bPhi & Q_A \bLambda & = \diff'_{\bA} \bLambda - \bmu_\C + \bB\\
	Q_A \bX & = \diff'_{\bA} \bX - \bPsi_{\bY} \qquad & Q_A \bPsi_{\bX} & = \diff'_{\bA} \bPsi_{\bX} - \bY \bPhi\\
	Q_A \bY & = \diff'_{\bA} \bY + \bPsi_{\bX} \qquad & Q_A \bPsi_{\bY} & = \diff'_{\bA} \bPsi_{\bY} - \bPhi \bX\\
\end{aligned}\,,
\ee
As expected, the deformation exactly matches $\partial_\varepsilon$ on the combined superfields $\CX, \CY, \CA, \CB$.

The (modified) stress tensor for the deformed theory is given by
\be
	\bT = -\bB \pd_z \bA + \bLambda \pd_z \bPhi + \tfrac{1}{2}\big(\bPsi_{\bX} \pd_z \bX + \bPsi_{\bY} \pd_z \bY - \bX \pd_z \bPsi_{\bX} - \bY \pd_z \bPsi_{\bY}\big),
\ee
from which it follows that $Q_A \bT = \diff \bT$ and, moreover, $\bS = \tfrac{1}{2}(\bY \pd_z \bX-\bX \pd_z \bY) - \bLambda \pd_z \bA$ solves Eq. \eqref{eq:QST} so that the theory is perturbatively topological. Unlike the previous examples, the $A$-twist of this $\CN=4$ gauge theory receives non-perturbative corrections in the form of monopole operators. Although we expect the full, non-perturbative theory to remain topological, it would be interesting to see how $\bS, \bT$ are modified once monopole operators are taken into account.

\subsubsection{Costello-Gaiotto boundary algebra}
\label{sec:bdySYMA}
We now move to the boundary algebra of \cite{CostelloGaiotto} for the $A$ twist of super Yang-Mills. Just as with the symplectic boson VOA for free hypermultiplets, this VOA plays a central role in the bulk 3d TQFT; a suitably defined category of its modules serves as a model for the category of bulk line operators, and its conformal blocks are related to state spaces of the bulk theory \cite{CostelloGaiotto, CostelloCreutzigGaiotto, GaiottoTwisted}.

The boundary conditions used by \cite{CostelloGaiotto} and \cite{BLS} are a (deformation of) $\CN=(0,4)$ Neumann boundary conditions, \ie\, we impose Neumann boundary conditions on the gauge fields and hypermultiplet scalars, extending to the remainders of their multiplets in a way compatible with $\CN=(0,4)$ supersymmetry. In terms of $\CN=2$ multiplets, we choose $\CN=(0,2)$ Neumann boundary conditions for the vector multiplet, $\CN=(0,2)$ Dirichlet boundary conditions for the adjoint chiral multiplet, and $\CN=(0,2)$ Neumann boundary conditions for the $T^*V$ chiral multiplets.

Importantly, there are potential gauge anomalies induced by this choice of boundary conditions that must be remedied by boundary degrees of freedom. Working in the conventions of \cite{CostelloDimofteGaiotto-boundary, DimofteGaiottoPaquette}, we find that the boundary conditions on the bulk degrees of freedom contribute $2h - T_{V}$ to the gauge anomaly, where $h$ is the dual Coexeter number of $G$ and $T_V$ is the quadratic index of $V$. Assuming the representation $V$ is sufficiently large, \ie\, $T_{V} \geq 2h$, we can cancel this gauge anomaly with boundary Fermi multiplets%
\footnote{This is always true for $G = \C^\times$ as $h = 0$ and $T_V = n^2$ for a representation of charge $n$. For $G = SL(N)$, which has $h = N$, with $K$ fundamental hypermultiplets, which each have $T_{{\rm fund}} = 1$, this constraint says that $K \geq 2N$. If the representation isn't sufficiently large we could instead choose boundary chiral multiplets to cancel the anomaly.} %
transforming in a representation $M$ of $G$, which contributes $T_M$ to the boundary anomaly, such that $T_M = T_{V}-2h$.

The boundary Fermi multiplets can be encoded in terms of twisted superfields $\bGamma, \wt{\bGamma}$, and the full boundary conditions on the bulk fields are the following.
\begin{itemize}
	\item Neumann boundary conditions for the $\CN = 2$ vector multiplet ($\bB|_{\pd} =  \bmu_\pd$)
	\item Dirichlet boundary conditions for the $\mathfrak{g}$ chiral multiplet ($\bPhi|_{\pd} = 0$)
	\item Neumann boundary conditions for the $V$ and $V^*$ chiral multiplets ($\bPsi_{\bX}|_{\pd}, \bPsi_{\bY}|_{\pd} = 0$)
\end{itemize}
In the above, $\bmu_\pd$ is the moment map for the $\mathfrak{g}$ action on $T^*M$. The choice of Dirichlet boundary conditions for $\bPhi$ ensures that the superpotential $\bW = -\bY \bPhi \bX$ vanishes at the boundary. Moreover, the Neumann boundary conditions on the vector multiplets ensure that there are no boundary monopole operators and thus it suffices to consider a perturbative analysis.

The Feynman diagram analysis of these boundary OPEs is nearly identical to what we saw in Section \ref{sec:bdyFHA}. The hypermultiplet part of the $HT$-twisted theory has the same propagators as above; the propagators for the vector multiplet fields take the same analytic form as above and will be represented by the following oriented edges.
\begin{figure}[H]
	\centering
	\begin{tikzpicture}
		\draw (1.9,0) node {$\bLambda_b$};
		\draw (-1.9,0) node {$\bPhi^a$};
		\draw[middlearrow={latex}, thick] (-1.41,0) -- (1.41,0);
	\end{tikzpicture}
	\qquad
	\begin{tikzpicture}
		\draw (5+1.9,0) node {$\bB_b$};
		\draw (5-1.9,0) node {$\bA^a$};
		\draw[middlearrow={latex}, thick] (5-1.41,0) -- (5+1.41,0);
	\end{tikzpicture}
\end{figure}
\noindent The $HT$-twisted theory has five trivalent interaction vertices coming from the gauge interactions and superpotential. The deformation to the $A$ twist induces two additional bivalent interaction vertices. Note that the highly restrictive structure of these interaction vertices and propagators indicates that this theory is 1-loop exact when we work in holomorphic gauge \cite{GwilliamWilliams}. 

The boundary OPEs then come from the following diagrams, which are computed just as in Section \ref{sec:bdyFHA}.
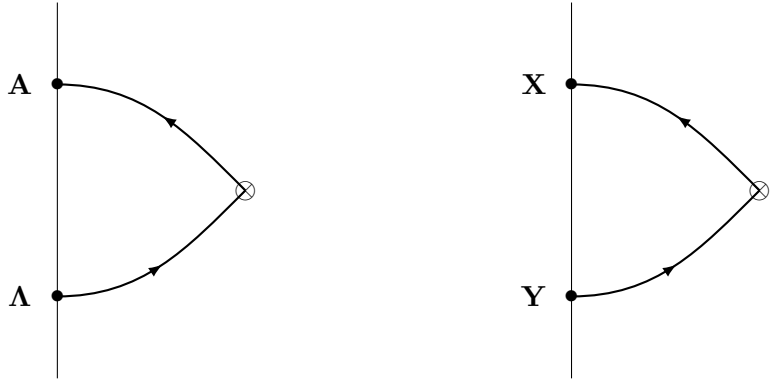
\begin{figure}[H]
	\centering
	\begin{tikzpicture}
		\node (int) at (0,0) {$\otimes$};
		\node (A) at (-3,1.41) {$\bA$};
		\node (bA) at (-2.5,1.41) {$\bullet$};
		\node (L) at (-3,-1.41) {$\bLambda$};
		\node (bL) at (-2.5,-1.41) {$\bullet$};
		\draw[middlearrow={latex}, thick] (bL.center) to [in=225, out=0] (int.center);
		\draw[middlearrow={latex}, thick] (int.center) to [in=0, out=135] (bA.center);
		\draw (-2.5,2.5) -- (-2.5,-2.5);
	\end{tikzpicture}
	\hspace{3cm}
	\begin{tikzpicture}
		\node (int) at (0,0) {$\otimes$};
		\node (X) at (-3,1.41) {$\bX$};
		\node (bX) at (-2.5,1.41) {$\bullet$};
		\node (Y) at (-3,-1.41) {$\bY$};
		\node (bY) at (-2.5,-1.41) {$\bullet$};
		\draw[middlearrow={latex}, thick] (bY.center) to [in=225, out=0] (int.center);
		\draw[middlearrow={latex}, thick] (int.center) to [in=0, out=135] (bX.center);
		\draw (-2.5,2.5) -- (-2.5,-2.5);
	\end{tikzpicture}
\end{figure}
\noindent The other interaction vertices, namely the trivalent vertices coming from coupling the gauge field to matter fields and the superpotential $-\bY \bPhi \bX$, do not contribute to singular terms in boundary OPEs. Thus, the boundary OPEs are given (up to a numerical prefactor) by
\be
\label{eq:SYMAope}
c^a(z) \lambda_b(w) \sim \frac{\delta^a{}_b}{z-w} \qquad X^n(z) Y_m(w) \sim \frac{\delta^n{}_m}{z-w}.
\ee
If $\gamma^\alpha, \wt{\gamma}_\beta$ are the lowest components of the boundary Fermi multiplets, they similarly have an OPE given by
\be
\gamma^\alpha(z) \wt{\gamma}_\beta(w) \sim \frac{\delta^\alpha{}_\beta}{z-w}.
\ee

Finally, there is a residual action of $Q_A$ on the boundary values of the fields:
\be
\label{eq:YMAQ}
\begin{aligned}
	Q_A c & = c^2 \qquad & Q_A \lambda & = -J\\
	Q_A X & = c \cdot X \qquad & Q_A Y & = c \cdot Y\\
	Q_A \gamma & = c \cdot \gamma \qquad & Q_A \wt{\gamma} & = c \cdot \wt{\gamma}\\
\end{aligned}\,,
\ee
where $J$ is the current generating the $\fg$ action; in components it reads
\be
J_a = f^c{}_{ab} \norm{\lambda_c c^b} + (\tau_a)^m{}_n \norm{Y_m X^n} + (\sigma_a)^i{}_j \norm{\wt{\gamma}_i \gamma^j}\,.
\ee
We therefore find that the boundary algebra is exactly the $G$-BRST reduction of $T^*V$-valued symplectic bosons (generated by $X,Y$) times $M$-valued complex fermions (generated by $\gamma, \wt{\gamma}$), exactly reproducing the boundary algebra of \cite{CostelloGaiotto}.

We note that the role of the $c$-ghost is to impose gauge invariance of boundary local operators in a derived fashion, \ie\, in a way compatible with taking the $A$ twist. As discussed in \cite{CostelloDimofteGaiotto-boundary}, this must be done with care. In particular, when the compact gauge group $G_c$ is semisimple, and correspondingly $G$ is reductive, the operation of taking $G$ invariants does not need to be derived (it's an exact functor) and hence taking $G$-invariants does not involve a corresponding ``$c$-ghost." In the present context, we are taking derived invariants for the group $G[\![z]\!]$ of formal series. The group $G[\![z]\!]$ schematically takes the form
\be
G[\![z]\!] \cong G \ltimes (1 + z \mathfrak{g}[\![z]\!])\,,
\ee
from which we see that the $z$-independent part of the group is reductive and hence we should not include the corresponding mode of $c(z)$. Instead, the process of taking derived $G[\![z]\!]$-invariants corresponds to restricting to $G$-invariants of the algebra generated by the bottom components $X, Y, \gamma, \wt{\gamma}, \lambda, \pd c$ and their derivatives, and then take cohomology with respect to $Q_A$.

\subsubsection{State spaces, local operators, and line operators}
\label{sec:hilbSYMA}
The geometric quantizations of the various space of solutions to the equations of motion of the above $A$-twisted theory is nearly the same as what appeared in Section \ref{sec:hilbFHA}. Just as with a free hypermultiplet, this geometric quantization on a Riemann surface $\Sigma$ is expected to reproduce the conformal blocks for the BRST reduction discussed in the previous subsection. Similarly, we can derive the category of line operators $\CC_A$ by considering the space of solutions on the formal punctured disk $\D^\times$.

The full, non-perturbative analysis is aided by collecting the twisted superfields as in Eq. \eqref{eq:combinedhypers} and Eq. \eqref{eq:combinedvect}, and will be nearly identical to Section \ref{sec:hilbFHA}. Prior to the deformation to the $A$ twist, the space of solutions to the equations of motion on a Riemann surface $\Sigma$ is given by the space of sections%
\be
\label{eq:EOMYMHT}
	\Sect(\Sigma_\opd \times \C[-1], T^*(V/G) \otimes K^{1/2}_\Sigma) \cong T[1]\Sect(\Sigma_\opd, T^*(V/G) \otimes K^{1/2}_\Sigma).
\ee
Just as with free hypermultiplets, deforming to the $A$ twist corresponds to adds the (holomorphic) de Rham differential to $Q_{HT}$, which deforms the above mapping space to%
\be
\label{eq:EOMYMA}
\begin{aligned}
	T[1]\Sect(\Sigma_\opd, T^*(V/G) \otimes K^{1/2}_\Sigma) \quad \rightsquigarrow \quad  & \Sect(\Sigma_\opd, T^*(V/G) \otimes K^{1/2}_\Sigma)_\dR\\
	& \quad \cong T^*\Sect(\Sigma_\opd, V/G \otimes K^{1/2}_\Sigma)_\dR\,.
\end{aligned}
\ee
A natural geometric quantization of this ($0$-shifted) cotangent bundle is again given by the (Borel-Moore) homology of the base
\be
\CH_A(\Sigma) \cong H_\bullet(\Sect(\Sigma_\opd, V/G \otimes K^{1/2}_\Sigma))\,.
\ee

From this analysis, we can extract the recent description of local operators in the $A$ twist due to Braverman-Finkelberg-Nakajima \cite{BFNII}; unlike the case of a free hypermultiplet, the $A$ twist of super Yang-Mills admits both perturbative and non-perturbative contributions to the algebra of local operators. The perturbative analysis is nearly identical to that of Section \ref{sec:hilbFHA} (one finds $G$-invariant polynomials in the zero-mode of the complex scalar $\phi$), so we only consider the full non-perturbative analysis here. We assume the surface is a raviolo $\Sigma = \D \cup_{\D^\times} \D$. We can then identify $\Sect(\Sigma_\opd, V/G \otimes K^{1/2}_\Sigma)$ with the space of holomorphic $G$-bundles on $\D \cup_{\D^\times} \D$, together with a section of an associated $V$-spinor-bundle. The data of such a bundle corresponds to a pair of holomorphic bundles $E, E'$ on $\D$, together with an identification $\sigma: E|_{\D^\times} \overset{\rm isom.}{\to} E'|_{\D^\times}$ over $\D^\times$. In addition, we have holomorphic sections $X(z)$ and $X'(z)$ of $(V \times_{G} E) \otimes K^{1/2}_\D$ and $(V \times_{G} E') \otimes K^{1/2}_\D$, taken to one another by the identification over $\D^\times$: $X'|_{D^*} = \sigma(X|_{\D^\times})$.

After trivializing the bundles, the identification $\sigma$ is simply an element of the loop group $g \in G(\!(z)\!)$. Similarly, the sections $X,X'$ are simply elements of $V[\![z]\!]$ that satisfy $X' = g X$. We can then identify the space of sections $\Sect(\Sigma_\opd, V/G \otimes K^{1/2}_\Sigma)$ with the preimage of $V[\![z]\!]$ under the map $m: G(\!(z)\!) \times V[\![z]\!] \to V(\!(z)\!), (g,X) \mapsto g X$, up to change of trivialization on the two disks, \ie\, up to acting with $G[\![z]\!] \times G[\![z]\!]$ via $(h',h).(g,X) = (h' g h{}^{-1}, h X)$
\be
\Sect((\D \cup_{\D^\times} \D)_\opd, V/G \otimes K^{1/2}_{\D \cup_{\D^\times} \D}) \cong G[\![z]\!] \backslash m^{-1}(V[\![z]\!])  /G[\![z]\!] \cong G[\![z]\!] \backslash \CR_{G,V}\,,
\ee
where $\CR_{G,V}$ is the ``space of triples" of \cite{BFNII}.

In contrast to the case of a free hypermultiplet, which had a contractible space of triples, the above space of sections for general gauge theories has highly non-trivial topology and geometry, as it would need to accommodate for monopole operators. Nonetheless, we claim that the algebra of local operators is realized as its the Borel-Moore homology%
\footnote{We note that it is important here to redefine the naive cohomological grading by the Coulomb-branch flavor symmetry to arrive at this answer. This choice of $R$-symmetry is precisely a choice of dimension theory on the above space of triples, cf. \cite[Section 1.3]{dmodules}. This allows for the definition the necessary semi-infinite cohomology used in defining the Coulomb branch.} %
\be
\label{eq:opsSYMA}
	\CH_A(\D \cup_{\D^\times} \D) \cong H_\bullet(G[\![z]\!] \backslash \CR_{G,V}) \cong H_\bullet^{G[\![z]\!]}(\CR_{G,V})\,.
\ee
This exactly reproduces the vector space of local operators in 3d $\CN = 4$ $G$ gauge theory with $T^*V$ hypermultiplet matter. The algebra structure on this vector space is induced by convolution as defined in \cite{BFNII}. 

A similar analysis goes through when considering line operators. The space of solutions to the equations of motion on the punctured disk $\Sigma = \D^\times$ is $T^*[1] \Sect(\D^\times_\opd, V/G \otimes K^{1/2}_{\D^\times})_\dR$. Just as with the free hypermultiplet, we choose a polarization induced by the base of the cotangent bundle to arrive at the category of line operators in the $A$-twist:
\be
	\CC_A = \textrm{QCoh}(\CL (V/G)_{\dR}) \cong D(\CL(V/G))
\ee
We have once again trivialized $K^{1/2}$ to identify $\Sect(\D^\times_\opd, V/G \otimes K^{1/2}_{\D^\times})$ with the algebraic loop space $\CL(V/G)$.

\subsubsection{Descent}
With the $A$-twisted action given above, we can again perform the descent analysis of Section \ref{sec:descent}. We will work perturbatively and assume the gauge group is abelian, as this captures the main features of interest.

The descent analysis on the hypermultiplets is the same as it was in Section \ref{sec:FHA}, so it suffices to consider a free $\CN = 4$ vector multiplet. The variations given in Eq. \eqref{eq:QSYM} and Eq. \eqref{eq:QASYM} imply that
\be
	(\phi^{(1)})^{\langle 1 \rangle} = -\pd_z c^{(1)} \wedge \diff z \qquad (B^{(1)})^{\langle 1 \rangle} = -\pd_z \lambda^{(1)} \wedge \diff z.
\ee
Let $\widetilde{\gamma}$ denote the (complexified) dual photon. It is related to $B$ by $\pd_z \widetilde{\gamma} \sim B$, \cf\, Section 3.4 of \cite{CostelloDimofteGaiotto-boundary}, whence
\be
\begin{aligned}
	\{\!\{\phi, \widetilde{\gamma}\}\!\} & = \{\!\{\pd_z c, \widetilde{\gamma}\}\!\}_{HT} = \{\!\{c, B\}\!\}_{HT} = 1\\
	\{\!\{\widetilde{\gamma}, \phi\}\!\} & = -\{\!\{\lambda, \phi\}\!\}_{HT} = -1\\
\end{aligned}\,.
\ee
Putting this together with the results of Section \ref{sec:FHA}, we get the expected result for the topological descent brackets of the $A$ twist super Yang-Mills \cite{descent}.

\subsubsection{Deformations induced by flavor symmetries}
\label{sec:SYMAdef}
We will end our analysis of the $A$ twist by discussing deformations of the above theory by background fields for flavor symmetries. The (complexified) flavor symmetries admitted by the above class of super Yang-Mills theories takes the form $G_H \times G_C$. The group $G_H$ is the group of (holomorphic symplectic) symmetries of the Higgs branch $\CM_H$; concretely, it is the normalizer of $G$ in ${\rm Sp}(T^*V)$, modulo the adjoint action of $G$. Similarly, the group $G_C$ is the group of (holomorphic symplectic) symmetries of the Coulomb branch $\CM_C$. Unlike $G_H$, only a maximal torus of $G_C$ is visible in the UV and is identified with the topological symmetry corresponding to abelian factors of the gauge group $G$.

We can turn on background fields coupling to the $G_H \times G_C$ flavor symmetry just as we did for the free hypermultiplet. Vector multiplets couple to $G_H$ flavor symmetries and thus we can deform the $A$-twisted theory by, \eg\,, a background holomorphic $G_H$-bundle. This is identical to what we saw with an $A$ twisted free hypermultiplet in Section \ref{sec:FHAdef}, so we focus on the $G_C$ flavor symmetry.

Unlike $G_H$, the $G_C$ flavor symmetry couples to \emph{twisted} vector multiplets. We thus expect to be able to deform the $A$-twisted theory by background \emph{flat} $G_C$-bundles. It is relatively straightforward to introduce background twisted superfields for the maximal torus $T_C \subseteq G_C$ that is visible in the UV. The abelian flat connection $\widetilde{\CA}$ is expressed in terms of the twisted superfields $\wt{\bA} \in \bOmega^{1,(0)}~\otimes~\mathfrak{t}_C[1]$ and $\wt{\bPhi} \in \bOmega^{0,(1)}\otimes \mathfrak{t}_C$. The coupling between this twisted vector multiplet and a usual vector multiplet is given by introducing a bilinear superpotential and Chern-Simons coupling \cite{KSmirrorsym}:
\be
\label{eq:SYMAdefaction}
S_A \rightsquigarrow S_A + \int\Tr(\bA) \pd \wt{\bA} + \Tr(\bPhi) \wt{\bPhi}\,.
\ee
As compared to the above, this action solves the classical master equation if 
\be
\diff' \wt{\bA} = 0 \qquad \diff' \wt{\bPhi} - \pd \wt{\bA} = 0\,,
\ee
\ie\, if $\wt{\bA}, \wt{\bPhi}$ actually come from a complex, abelian flat connection.

The effect this abelian flat connection has on the boundary algebra of Section \ref{sec:bdySYMA} is more transparent if we work in a holomorphic gauge $\wt{\bA} = 0, \wt{\bPhi} = \wt{A}(z) \diff z$: the deformation doesn't change any of the OPEs of the fundamental fields in the boundary VOA described in Section \ref{sec:bdySYMA} but does change the action of the differential $Q_A$
\be
Q_A \Tr(\lambda) =  -\Tr(J) \quad \rightsquigarrow \quad Q_A \Tr(\lambda) =  - \Tr(J) + \wt{A}(z)\,.
\ee
This change in the action of $Q_A$ results in a deformed OPEs of cohomology classes.

As an example of this phenomenon, consider the case of $\CN=4$ $U(1)$ gauge theory with $1$ hypermultiplet. The $\CN=(0,4)$ boundary condition involves introducing a (charge 1) boundary Fermi multiplet. The deformed boundary algebra is realized as the cohomology of the $U(1)$-invariants of the VOA generated by $\pd c, \lambda, X, Y, \gamma, \wt{\gamma}$ with respect to the differential
\be
\begin{aligned}
	Q_A c & = 0 \qquad & Q_A \lambda & = -J + \wt{A}\\
	Q_A X & = c X \qquad & Q_A Y & = -c Y\\
	Q_A \gamma & = c \gamma \qquad & Q_A \wt{\gamma} & = -c \wt{\gamma}\\
\end{aligned}\,,
\ee
where $J = \norm{YX} + \norm{\wt{\gamma}\gamma}$. We find that the cohomology is generated by the fermionic currents $\psi_1 = \wt{\gamma} X$ and $\psi_2 = Y \gamma$, with OPE given by
\be
\begin{aligned}
	\psi_1(z) \psi_2(w) & \sim \frac{1}{(z-w)^2} + \frac{1}{z-w} (\norm{YX}(w) + \norm{\wt{\gamma} \gamma}(w))\\
	& \qquad = \frac{1}{(z-w)^2} + \frac{\wt{A}(w)}{z-w} + Q_A(...)\,.
\end{aligned}
\ee
As expected, this is exactly mirror to the deformation of a $B$-twisted hypermultiplet by an abelian flat connection described in Section \ref{sec:FHB}.

\subsection{$B$ twist}
\label{sec:SYMB}
We now turn to the $B$ twist of the super Yang-Mills. The twisting takes the same form as in Section \ref{sec:FHB}. The new twisted spin and cohomological gradings are given below.

\begin{table}[H]
	\centering
	\begin{tabular}{c|c|c|c|c|c|c|c|c}
		& $\bA$   & $\bB$  & $\bPhi$ & $\bLambda$ & $\bX$ & $\bPsi_{\bX}$ & $\bY$ & $\bPsi_{\bY}$ \\ \hline
		$J_B$ & $0$ & $1$ & $1$ & $0$ & $0$   & $1$    & $0$ & $1$\\
		$C_B$ & $1$ & $0$ & $0$ & $1$ & $0$   & $1$    & $2$ & $-1$  
	\end{tabular}
	\caption{Twisted spin $J_B$ and cohomological degree $C_B$ of the twisted superfields in the topological $B$ twist of super Yang-Mills.}
	\label{table:Bspincoho}
\end{table}

Again, we expect that the deformations for the hypermultiplets and vector multiplets are decoupled from one another, so it suffices to figure out how the vector multiplet transforms under $\oQ^2_-$. We find that $\oQ^2_- (A_t -i \sigma) = \oQ^2_- \lambda_+ = 0$, $\oQ^2_- (F_{zt} + i D_z \sigma) \sim \pd \lambda_+$ and $\oQ^2_- \varphi \sim \pd \lambda_-$. Using the fact that the ($z$ covariant derivative of the) $c$ ghost is cohomologous to $\lambda_-$ and $(F_{zt} + i D_z \sigma)$ is $B$, the deformation of the action is given by
\be
\label{eq:BtwistedactionYM}
S_B = S + \int\bY \pd \bX - \bLambda \pd \bA\,,
\ee
corresponding to the following action of $Q_B$:
\be
\begin{aligned}
	\label{eq:QBYM}
	Q_B \bA & = F'(\bA) \qquad & Q_B \bB & = \diff'_{\bA} \bB - \bmu - \pd \bLambda\\
	Q_B \bPhi & = \diff'_{\bA} \bPhi - \pd \bA & Q_B \bLambda & = \diff'_{\bA} \bLambda - \bmu_\C\\
	Q_B \bX & = \diff'_{\bA} \bX \qquad & Q_B \bPsi_{\bX} & = \diff'_{\bA} \bPsi_{\bX} - \bY \bPhi - \pd \bY\\
	Q_B \bY & = \diff'_{\bA} \bY \qquad & Q_B \bPsi_{\bY} & = \diff'_{\bA} \bPsi_{\bY} - \bPhi \bX + \pd \bX\\
\end{aligned}
\ee

The modified stress tensor $\bT$ for this $B$-twisted theory is given by
\be
	\bT = - \bB \pd_z \bA - \bPhi \pd_z \bLambda + \bPsi_{\bX} \pd_z \bX + \bPsi_{\bY} \pd_z \bY\,,
\ee
which indeed satisfies $Q_B \bT = \diff \bT$. Moreover, it is straightforward to check that we can solve Eq. \eqref{eq:QST} with ${\bS = \bB \iota_{\pd_z} \bPhi - \bPsi_{\bX} \iota_{\pd_z} \bPsi_{\bY}}$, so that the analysis of Section \ref{sec:Qstress} implies this theory is perturbatively topological. We expect this extends to the full quantum theory as there are no non-perturbative corrections in the $B$-twist.

We note that the action again takes a remarkably simple form
\be
\label{eq:BtwistedactionSYM2}
S_B = \int \CB F(\CA) + \CY \diff_\CA \CX\,,
\ee
agreeing with the AKSZ theory based on the mapping space $\Maps(\R^3_\dR, T^*[2] (V/G))$ \cite{ESWtax}, as expected; this is also known as ``BF-Rozansky-Witten theory" \cite{KQZ}. In fact, if we redefine the $R$-charge so that $\CX,\CY$ have cohomological degree 1, we can view this as a Chern-Simons theory based on the Lie superalgebra $\widehat{\mathfrak{g}} = T^*\mathfrak{g} \oplus \Pi (T^* V)$ \cite{CostelloGaiotto}, with the following non-trivial brackets:
\be
\label{eq:SYMBliealg}
\begin{alignedat}{3}\relax
	[T_a, T_b] & = f^c{}_{ab} T_c &&&  [T_a, S^b] & = f^b{}_{ac} S^c\\
	[T_a, \theta_m] & = \theta_n (\tau_a)^n{}_m &&&  [T_a, \otheta^n] &= -(\tau_a)^n{}_m \otheta^m\\
	&& \hspace{-1.5cm} \{\theta_m, \otheta^n\} &= (\tau_a)^n{}_m S^a \hspace{-1.5cm} &&\\
\end{alignedat}
\ee
In these formulae, the $T_a$ are basis vectors for $\mathfrak{g}$, with $f^c{}_{ab}$ the structure constants in that basis, $S^b$ are dual basis vectors for $\mathfrak{g}^*$, $\theta_m$ and $\otheta^m$ are dual bases of $\Pi V$ and $\Pi V^*$, with $(\tau_a)^n{}_m$ the matrices representing the $\mathfrak{g}$ action on $V$ and $V^*$.

\subsubsection{Costello-Gaiotto boundary algebra}
\label{sec:bdySYMB}
First consider the boundary algebra of \cite{CostelloGaiotto} for the $B$ twist of super Yang-Mills. The appropriate boundary condition is a (deformation of a) $\CN=(0,4)$ Dirichlet boundary condition. In particular, we impose Dirichlet boundary on the gauge fields and require that the hypermultiplet scalars vanish at the boundary, extending by $\CN=(0,4)$ supersymmetry. In terms of the $B$-twisted superfields, this boundary conditions is given as follows.
\begin{itemize}
	\item Dirichlet boundary conditions for the $\CN = 2$ vector multiplet ($\bA|_{\pd} = 0$)
	\item Neumann boundary conditions for the $\mathfrak{g}$ chiral multiplet ($\bLambda|_{\pd} = 0$)
	\item Dirichlet boundary conditions for the $V$ and $V^*$ chiral multiplets ($\bX|_{\pd}, \bY|_{\pd} = 0$)
\end{itemize}
These boundary conditions ensure that the superpotential vanishes as the boundary. Unlike the $A$ twist analysis in Section \ref{sec:bdySYMA}, the choice of Dirichlet boundary conditions for the vector multiplet means we do not need to include boundary degrees of freedom to cancel gauge anomalies. On the other hand, there are non-perturbative local operators on the boundary. We will leave the full, non-perturbative analysis of boundary monopoles to future work and only consider the perturbative part of the boundary algebra.

The propagators are essentially the same as those discussed in Section \ref{sec:bdyFHB} and \ref{sec:bdySYMA}. In addition to the trivalent vertices already present in the $HT$-twisted theory, there are two bivalent vertices coming from the $B$ twist deformation. Again, the highly restrictive structure of these interaction vertices and propagators implies that this theory is 1-loop exact.

The only fields that survive cohomology at the boundary are the bottom components $B, \phi, \psi_X, \psi_Y$. Moreover, once the boundary conditions are imposed, the action of $Q_B$ is trivial on these fields. The OPEs between operators built from these fields can be obtained in a manner similar to \cite{GwilliamWilliams} and \cite{CostelloDimofteGaiotto-boundary}. For example, a $BB$ OPE should be induced as follows. First, there should be a first-order pole proportional to $B$ coming from the trivalent vertex of the gauge covariant derivative. Additionally, there should a second-order pole coming from the (four) 1-loop diagrams.
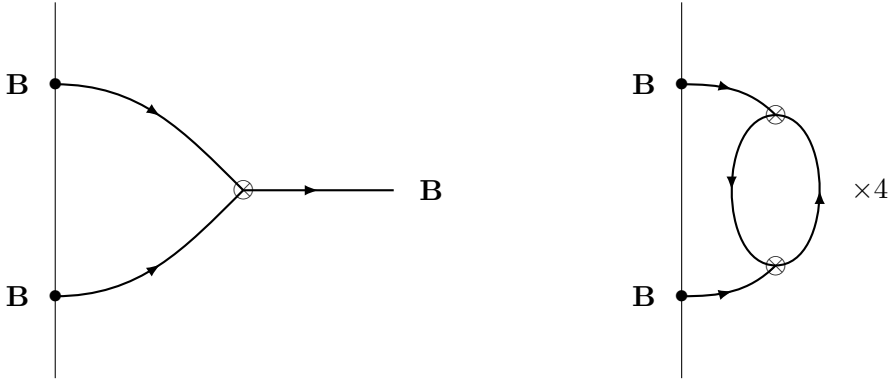
\begin{figure}[H]
	\centering
	\begin{tikzpicture}
		\node (int) at (0,0) {$\otimes$};
		\node (B1) at (-3,1.41) {$\bB$};
		\node (bB1) at (-2.5,1.41) {$\bullet$};
		\node (B2) at (-3,-1.41) {$\bB$};
		\node (bB2) at (-2.5,-1.41) {$\bullet$};
		\node (B3) at (2.5,0) {$\bB$};
		\node (bB3) at (2,0) {};
		
		\draw[middlearrow={latex}, thick] (bB1.center) to [in=135, out=0] (int.center);
		\draw[middlearrow={latex}, thick] (bB2.center) to [in=225, out=0] (int.center);
		\draw[middlearrow={latex}, thick] (int.center) to [in=180, out=0] (bB3.center);
		\draw (-2.5,2.5) -- (-2.5,-2.5);
	\end{tikzpicture}
	\hspace{2cm}
	\begin{tikzpicture}
		\node (int1) at (-1.25,1) {$\otimes$};
		\node (int2) at (-1.25,-1) {$\otimes$};
		\node (B1) at (-3,1.41) {$\bB$};
		\node (bB1) at (-2.5,1.41) {$\bullet$};
		\node (B2) at (-3,-1.41) {$\bB$};
		\node (bB2) at (-2.5,-1.41) {$\bullet$};
		
		\node at (0,0) {$\times 4$};
		
		\draw[middlearrow={latex}, thick] (bB1.center) to [in=135, out=0] (int1.center);
		\draw[middlearrow={latex}, thick] (bB2.center) to [in=225, out=0] (int2.center);
		
		\draw[middlearrow={latex}, thick] (int1.center) to [in=180, out=180] (int2.center);
		
		\draw[middlearrow={latex}, thick] (int2.center) to [in=0, out=0] (int1.center);
		
		\draw (-2.5,2.5) -- (-2.5,-2.5);
	\end{tikzpicture}
\end{figure}
\noindent Loops with $\bB/\bA$ and $\bPhi/\bLambda$ each contribute a term proportional to $-h\kappa_{ab}$, while the loops with $\bX/\bPsi_{\bX}$ and $\bY/\bPsi_{\bY}$ each contribute terms proportional to $\tfrac{1}{2} T_V \kappa_{ab}$, where $\kappa_{ab}$ is the Killing form on $G$. The remaining OPEs admit a similar analysis. All together, we expect that the OPEs take the following form (up to overall numerical factors):
\begin{gather}
	\label{eq:opeSYMB}
	B_a(z) B_b(w) \sim  \frac{(T_V - 2h)\kappa_{ab}}{(z-w)^2} + \frac{f^c{}_{ab}}{z-w} B_c(w) \qquad \psi_Y{}^n(z) \psi_X{}_m(w) \sim \frac{\delta^n{}_m}{(z-w)^2} + \frac{(\tau_a)^n{}_m}{z-w} \phi^a(w) \nonumber \\
	B_a(z) \phi^b(w) \sim  -\frac{\delta^b{}_a}{(z-w)^2} + \frac{f^b{}_{ac}}{z-w}\phi^c(w)\\
	B_a(z) \psi_Y{}^n(w) \sim -\frac{(\tau_a)^n{}_m}{z-w}\psi_Y{}^m(w) \qquad B_a(z) \psi_X{}_n(w) \sim \frac{ (\tau_a)^m{}_n}{z-w}\psi_X{}_m(w) \nonumber
\end{gather}
We arrive at the following description of the perturbative algebra: it is the affine VOA based on the Lie superalgebra $T^*\mathfrak{g} \oplus \Pi(T^*V)$. The central extension of the loop algebra used in defining the affine algebra comes from the Killing form on $\mathfrak{g}$ (proportional to the anomaly of the $G$ boundary flavor symmetry), the (negative of the) pairing of $\mathfrak{g}$ and $\mathfrak{g}^*$, and the symplectic form on $\Pi(T^* V)$. This result exactly matches the perturbative analysis of \cite{CostelloGaiotto}.

\subsubsection{State spaces, local operators, and line operators}
\label{sec:hilbSYMB}
With the above description of the $B$-twisted theory, we now turn to analyzing the state spaces of the theory on $\Sigma \times \R$.

We start with the full, non-perturbative analysis. Prior to deforming to the $B$ twist, we find that solutions to the equations of motion are given by points of
\be
\label{EOMSYMHT2}
\Maps(\Sigma_\opd \times \C^{0|1}, T^*[2](V/G)) \cong \Maps(\Sigma_\Dol, T^*[2](V/G))\,.
\ee
Just like a free hypermultiplet, when we combine the fields into $\CA, \CB, \CX, \CY$ as in Eq. \eqref{eq:combinedhypers} and \eqref{eq:combinedvect}, the $B$ twist deformation corresponds to including the additional differential $-\varepsilon \pd$, thus the deformation to the $B$ twist corresponds to deforming $\Sigma_\Dol$ to $\Sigma_\dR$
\be
\label{EOMSYMB}
\Maps(\Sigma_\Dol, T^*[2](V/G)) \rightsquigarrow \Maps(\Sigma_\dR, T^*[2](V/G)) \cong T^*\Maps(\Sigma_\dR, V/G)\,.
\ee

Geometric quantization of the above ($0$-shifted) cotangent bundle proceeds as before. We choose the polarization induced by the complex structure, leading to functions on the same mapping space appearing in Section \ref{sec:hilbFHB}.
\be
	\CH_B(\Sigma) \cong \C[\Maps(\Sigma_\opd, T^*[2](V/G))]\,.
\ee
This cannot be simplified further for general $\Sigma, V,G$%
\footnote{The result for the Hilbert space in Rozansky-Witten theory in terms of the sheaf cohomology of exterior powers of the tangent bundle derived in \cite{RW}, cf. Eq. \eqref{eq:hilbFHB}, need not hold in gauge theory, where one really has to contend with the Higgs branch as stack rather than simply the smooth hyperk\"{a}hler or holomorphic-symplectic manifold used in Rozansky-Witten theory.}, %
but the genus 0 case $\Sigma = S^2$ admits a simpler description:
\be
	\Maps(S^2_\opd, T^*[2]V/G) \cong T^*[2] V/G.
\ee
This analysis immediately tells us that the space of local operators of the theory is simply given by holomorphic functions on the (derived) Higgs branch $T^*[2](V/G)$
\be
	\CH_B(S^2) \cong \C[T^*[2](V/G)]\,.
\ee
A concrete model for the ring of functions is as $G$-invariant functions on $T^*[2]V \times \mathfrak{g}^*[1]$ subject to the differential
\be
\begin{aligned}
	& \qquad Q_B \lambda = -\mu_\C \\
	& Q_B X = 0 \qquad Q_B Y = 0\\
\end{aligned}\,.
\ee

Just as with the case of a free hypermultiplet, there are no non-perturbative local operators, \ie\, monopole operators, in the $B$ twist, as $Q_B$ cohomology localizes to flat connections. Indeed, the above answer for local operators can be obtained from a purely perturbative analysis, as we now show. We consider a spectral sequence whose first page comes from the $\diff$ part of $Q_B$; this removes all components of $\CA, \CB, \CX, \CY$ with non-trivial form degree as well as the dependence on insertion points of the bottom components. Naively, we are left with computing the following cohomology%
\be
\begin{aligned}
	\label{eq:pg2YMB}
	\delta_B c & = c^2 \qquad & \delta_B \lambda & =  c \cdot \lambda - \mu_\C\\
	\delta_B X & = c \cdot X & \qquad \delta_B Y & = c \cdot Y\\
\end{aligned}\,.
\ee
Of course, we should remove the zero-mode of the $c$ ghost and instead take $G$-invariants by hand. We are thus lead to the exact same conclusion as above: the algebra of local operators is given by functions on the Higgs branch $T^*[2](V/G)$.

To end this subsection, we derive the category of line operators via geometric quantization of the space of solutions to the equations of motion on the formal punctured disk $\Sigma = \D^\times$. As with our other examples, the space of solutions can be identified with a ($1$-shifted) cotangent bundle $\Maps(\D^\times_\dR, T^*[2]V/G) \cong T^*[1] \Maps(\D^\times_\dR, V/G)$. The category of line operators arising from geometric quantization with polarization induced by the base of this cotangent bundle is again quasi-coherent sheaves on the base
\be
	\CC_B = \textrm{QCoh}(\Maps(\D^\times_\dR, V/G))\,.
\ee
cf. \cite{linevortex, HilburnRaskin}.

\subsubsection{Descent}
Just as with the $A$ twist, we can use the deformation to the $B$ twist found above to perform the descent procedure of Section \ref{sec:descent}. Again, the computations for the hypermultiplets and vector multiplets decouple. For an $\CN = 4$ vector multiplet, we find 
\be
(c^{(1)})^{\langle 1 \rangle} = \phi^{(1)} \wedge \diff z \qquad (\lambda^{(1)})^{\langle 1 \rangle} = B^{(1)} \wedge \diff z
\ee
which imply that 
\be
\begin{aligned}
	\{\!\{c, \lambda\}\!\} = \{\!\{\phi, \lambda\}\!\}_{HT} = 1 \\
	\{\!\{\lambda, c\}\!\} = \{\!\{B, c\}\!\}_{HT} = 1\\
\end{aligned}.
\ee
This computation is somewhat misleading, as the zero mode of the $c$ ghost should not be included in the algebra of local operators (perturbative or otherwise) and so $\lambda$ trivially brackets with all other operators. Putting this together with the results of Section \ref{sec:FHB}, we have recovered the descent brackets for the $B$ twist of super Yang-Mills.

\subsubsection{Deformations induced by flavor symmetries}
\label{sec:SYMBdef}
As with the $A$ twist discussed in Section \ref{sec:SYMAdef}, we can turn on background fields that couple to the $G_H \times T_C$ flavor symmetry of the $B$ twist. Indeed, the deformations take the same form as in the $A$-twist. The only difference as compared to the $A$ twist is that we must exchange the types of bundles allowed: we can deform by complexified, flat $G_H$ connection, \cf\, Section \ref{sec:FHBdef}, and holomorphic $T_C$ bundles (or more generally monopole configurations). In particular, the above deformed action only satisfies the classical master equation if 
\be
\begin{aligned}
	\diff' \wt{\bA} + \wt{\bPhi} & = 0 & \qquad \diff' \wt{\bPhi} & = 0\\
	F'(\widehat{\bA}) & = 0 & \qquad \diff'_{\widehat{\bA}} \widehat{\bPhi} - \pd \widehat{\bA} & = 0\\
\end{aligned}\,.
\ee

Turning on a non-trivial background $T_C$ bundle doesn't affect the OPEs of the boundary algebra described in Section \ref{sec:bdySYMB}. Just as with background $G_H$ bundles in the $A$-twist, these backgrounds change will change the gluing rules of correlation functions of operators charged under $T_C$, \ie\, boundary monopole operators. The deformation by a background flat $G_H$ connection is much more dramatic. If we work in a holomorphic gauge where $\widehat{\bA} = 0$ and $\widehat{\bPhi} = \widehat{A}(z) \diff z$, the effect of deforming the theory is a straightforward generalization of that discussed in Section \ref{sec:FHBdef}. In particular, the boundary VOA discussed in Section \ref{sec:bdySYMB} is exactly the same with a slightly modified OPE for $\psi_X, \psi_Y$:
\be
	\psi_Y{}^n(z) \psi_{X m}(w) \sim \frac{\delta^n{}_m}{(z-w)^2} + \frac{\widehat{A}^n{}_m(w)}{z-w} + \frac{(\tau_a)^n{}_m}{z-w} \phi^a(w)\,.
\ee

\acknowledgments
We would like to thank Tudor Dimofte for his support during the preparation of this paper and his suggestion for investigating this problem. We would also like thank Kevin Costello, Thomas Creutzig, Justin Hilburn, Brian Williams, and Keyou Zeng for useful conversations during the development of this project. N.G. acknowledges support from the University of Washington and previous support from T. Dimofte's NSF CAREER grant DMS 1753077.


\bibliography{twisted}
\bibliographystyle{JHEP_TD}
	
\end{document}